\documentclass[preprint,eqsecnum,aps,showpacs]{revtex4}

\usepackage{epsfig}

\usepackage{amssymb}
\usepackage{amsfonts}
\usepackage{amsmath}

\newcommand{\ds}{\displaystyle}
\newcommand{\ts}{\textstyle}
\newcommand{\be}{\begin{equation}}
\newcommand{\ee}{\end{equation}}
\newcommand{\ba}{\begin{eqnarray}}
\newcommand{\ea}{\end{eqnarray}}
\newcommand{\eb}{\epsilon_{B}}
\newcommand{\ed}{\epsilon_{D}}

\begin{document}

\preprint{}
\title{Two-Component Cosmological Fluids with Gravitational Instabilities}
\author{R. M. Gailis}
\email{ralph.gailis@dsto.defence.gov.au}
\altaffiliation[Permanent address: ]{Defence Science and Technology
Organisation, 506 Lorimer St., Fishermans Bend, Victoria 3207, Australia}
\author{N. E. Frankel}
\email{n.frankel@physics.unimelb.edu.au}
\thanks{Corresponding author}
\affiliation{
School of Physics,
University of Melbourne,
Parkville, Victoria 3052,
Australia}

\begin{abstract}
A survey of linearized cosmological fluid equations with a number of
different matter components is made.
To begin with, the one-component case is reconsidered to illustrate some
important mathematical and physical points rarely discussed in the
literature.
The work of some previous studies of two-component systems are examined
and re-analyzed to point out some deficiencies of solutions, and further
solutions and physical interpretation are then presented.
This leads into a general two-component model with variable velocity
dispersion parameters and mass density fractions of each component.
The equations, applicable to both hot dark matter (HDM) and cold dark
matter (CDM) universes are solved in the long wavelength limit.
This region is of interest, because some modes in this range of
wavenumbers are Jeans unstable.
The mixture Jeans wavenumber of the two-component system is introduced
and interpreted, and the solutions are discussed, particularly in
comparison to analogous solutions previously derived for plasma modes.
This work is applicable to that region in the early Universe ($20 < z < 140$),
where large scale structure formation is thought to have occurred.
\end{abstract}


\maketitle

\section{Introduction}

The theory of structure formation in the Universe has become one of the
most popular and intensely studied topics in modern cosmology.
Throughout the past century there has been an accumulating volume of work 
on the analytic investigation of the cosmological structure formation
equations.
The various approaches include both fluid and kinetic theory formulations.
They principally consider the gravitational interaction of components of
the cosmological medium, though sometimes other forms of interaction such
as magnetic fields are also included (for some standard examples see
e.g. \cite{padman,peebles}).
The analysis of these equations has employed ever more diverse and 
complicated techniques and approximation schemes to model increasingly 
realistic physical situations.
This has been comprehensively supported and now superseded by large 
N-body simulations. 
The algorithms which govern these large numerical studies have grown 
progressively more refined and subtle, and are now producing very accurate
and realistic results, which can be directly compared with observations
(e.g.\cite{virgo}).

Despite the current trends in modern cosmological structure formation
theory, much can still be learned from relatively simple analytic models.
We consider such models, in the face of modern computing power, to analyze
at a fundamental level some of the basic physical processes which cause 
the clustering observed in the Universe.
This helps to isolate physical mechanisms difficult to discern numerically.
In this paper our interest will focus on the linearized cosmological fluid
equations.
These equations have been used to build up the components of the 
cosmological density perturbation power spectrum, and must be evolved 
through the various stages of cosmological evolution, and over a large 
range of physical scales.
There are several reasons for taking such an approach.
The equations may be solved numerically to give detailed power spectra
for the various cosmological models currently viable.
The power spectra may then be used as initial data to evolve the large 
N-body simulations, which are ultimately compared to observations.
The equations may also be used to build up a semi-quantitative picture of 
the evolution of the power spectrum.
This can show how the various sized perturbations scale with respect to 
the Friedmann expansion parameter $a$ during different epochs of the 
Universe, and give a direct insight into some of the fundamental physical 
processes operating to produce structure in the Universe (see e.g.
\cite{padman}).

The evolution of perturbation modes with wavelength greater than the 
Hubble radius may be studied through a relativistic formulation of the 
perturbation equations, whereas for modes with wavelengths smaller than 
this radius, a Newtonian formalism suffices.
There is a range of physical parameters and variety of differential equations
describing the evolution processes of density fluctuations in the early
Universe.
This involves such elements as equations of state for energy constituents,
and specification of the expansion parameter $a$ by the Friedmann cosmological
equations.
As a consequence, there is a large wealth of literature on this subject, 
and most of the currently important techniques and results have been 
collected in some well known textbooks \cite{padman,peebles,texts}.
Some of the relatively complicated systems of equations have also been
studied in the literature, and it is our goal to both review many of these 
studies, and to extend them in new directions, achieving some new unique
results.

In this paper we will only study the Newtonian limit of the linearized 
cosmological perturbation equations, valid for density fluctuations on 
scales well within the Hubble radius.
Our main concern is with some mathematically more complicated 
multi-component models, which although not usually considered in standard 
power spectrum analysis, have realistic and interesting physical meaning.
The primary concept of the Jeans gravitational instability \cite{jeans}
has been investigated in a static universe for multi-component models, to
reveal the more complicated structure of modes possible 
\cite{russians1,carvalho}.
This provides some interesting qualitative ideas about the possible
mechanisms for structure formation, but the lack of an expanding background
spacetime in the models leads to unrealistic solutions, exponential in
form.
The inclusion of cosmological expansion in the equations leads to the
more realistic power law and logarithmic solutions, familiar from the
standard power spectrum analysis.
Previous work in this area has focused both on some particular models
\cite{russians2}, and on a more general classification of the 
equations and solutions for a range of parameter values (some of them 
only of mathematical interest) and physical contexts 
\cite{fargion,haubold1,haubold2}.
Analytic solutions for some of the most general cases of the equations
considered above, which often have
significant physical interest, have not been achieved.
It is our aim here to rectify this situation and investigate
a system of equations modeling a two-component fluid in the matter 
dominated post-recombination era of an Einstein--deSitter universe.
One of these components consists of baryons, and the other some form of 
nonrelativistic dark matter particles.
Through this work we will amend what appear to be some errors in the
previous general studies of Haubold and Mathai \cite{haubold2}.

This paper makes a comprehensive study of the linear perturbation equations
for cosmological fluids with gravitational instabilities with application
to large scale structure formation.
For a historical perspective we note that Lifshitz \cite{lifshitz} concluded
that gravitational instability could not be responsible for the formation of
structure in the Universe.
The correct conclusion, that gravitational instability suffices, was pointed
out by Novikov \cite{novikov}.
As our work details the mathematical structure of the appropriate equations
describing cosmological structure formation, we note the nice series of papers
by Ratra and Peebles \cite{ratra} directed to understanding the applications
of special functions to the problem of gravitational instability in
cosmological models.
We further note the nice series by Buchert {\em et al} \cite{buchert}
concerning analytic results and their relevance to observational cosmology.

We will also make a comparison with work done in cosmological plasma
physics in an Einstein-deSitter background \cite{dettmann,plasma,gailis}.
This is interesting due to the mathematically very similar form of
fluid equations for both type of systems.
This similarity is largely due to the similarity of the electromagnetic
and gravitational forces.
In this paper we will analyze the long wavelength region of the solutions.
This corresponds to the Jeans unstable region of parameter space, and
requires use of Frobenius methods of expansion of the differential
equations.
In a follow-up paper \cite{paper2} we will investigate the short wavelength
region of the solutions, which will require a WKB approximation scheme to
be developed.

In these papers we take explictly the temperature relationship
$T \sim 1/a(t)^2$, where $a(t)$ is the radius of the Universe.
We give here the explanation why this is so.
Following standard textbook material in Padmanabhan \cite{padman} (as
summarized in equation (3.118)) and Peebles (1993, page 179) we find that the
baryons follow this relationship when
\[ 1 + z = 142( \Omega_b h^2 / 0.024)^{2/5}. \]
Here $\Omega_b$ has been scaled to the WMAP best fit value.
Thus for redshifts below $z \sim 140$ the baryon temperature drops as $1/a^2$
down to $z \sim 20$, where the Universe reionizes (probably in a patchy
fashion) and the $1/a^2$ scaling no longer holds.
This $20 < z < 140$ redshift region is important, as it is where early large
scale structure formation is thought to have occurred.

The paper is to be organized as follows.
In Section~2 we introduce the most general cosmological density perturbation
equations in the Newtonian approximation.
We review and classify previous work on these equations to put our current
work into context, showing what has been achieved, what needs amendment,
and where we will seek to expand current knowledge.
In Section~3 we revert to the one-component equations, to illustrate some
of the basic principles which will be important later in our analysis, and
to reveal some apparently new results.
This will enable us to begin to tackle the two-component problem in
Section~4.
In this section the CDM two-component
model in an expanding universe will be investigated.
This has ties with the previous work cited, and we will demonstrate the
limitations of the existing formalism here.
We present new results apparently overlooked in the work of Haubold and
Mathai \cite{haubold2}.
After this, we are ready to study the most general baryonic and dark matter
equations in Section~5, where we will consider the long wavelength 
approximation, applicable to either HDM or CDM.
This is followed by our conclusions in Section~6.

\section{A Classification of Cosmological Density Perturbation Equations}

As discussed in Section~1, there are a vast spectrum of equations 
describing cosmological density perturbations in different physical
regimes.
We begin directly with the linearized Newtonian approach.
The equations for an $n$-component system of nonrelativistic species is
derived in all the standard texts.
Given a density perturbation $\delta_{i}$ in the $i$-th component of the
mass density $\rho_{i}$:
\be
\delta_{i}({\bf r}, t) = \frac{\delta\rho_{i}}{\rho_{i}},
\ee
it may be decomposed into its Fourier plane wave modes with wave vector
${\bf k}$
\be
\delta_{i}({\bf r}, t) = \frac{1}{(2 \pi)^{3}} \int \delta_{{\bf k}i}(t)
    \exp(-i{\bf k}\cdot{\bf r}) d^{3}r.
\ee
Here ${\bf r}$ is the physical spatial coordinate, and $t$ is cosmic time.

To be able to solve the equations, the implicit time dependence of the 
physical variables needs to be removed.
We will adopt the convention that barred variables will denote comoving
quantities, independent of time.
Thus we define the comoving wave number $\bar{\bf k} = a{\bf k}$.
Using the Eulerian equations of motion describing a perfect
fluid, a set of coupled second order equations for the Fourier modes 
$\delta_{i}(t)$ (where we now drop the subscript ${\bf k}$) are achieved:
\be
\frac{d^{2}\delta_{i}}{dt^2} + 2\frac{\dot{a}}{a}\frac{d\delta_{i}}{dt}
    + \frac{v_{i}^{2} \bar{k}^{2}}{a^{2}} \delta_{i} = 
    4\pi G \sum_{i=1}^{n} \rho_{i}\delta_{i},\;\;\;\; i = 1, 2, \ldots n.
\label{general-cos-equs}
\ee
Overdots will denote derivatives with respect to $t$.
The above equations contain the sound velocity
\be
v_{i}^2 = \frac{dp_{i}}{d\rho_{i}} \propto \rho_{i}^{\gamma_{i}-1}.
\ee
The sound velocity depends on the equation of state of the medium, and
is in general dependent on time.
We have introduced the specific heat ratio $\gamma_i$, and assumed an
equation of state of the form $p_{i} \propto \rho_{i}^{\gamma_{i}}$.

The introduction of a sound velocity implicitly assumes that the fluids
involved are collisional.
This means that there are considerable interactions between the
particles comprising each matter component.
It is in fact generally assumed that dark matter is {\em collisionless},
in which case a fluid equation is not strictly correct.
The dark matter component would better be modeled by a distribution
function satisfying the Vlasov equation.
``Fluid-like'' equations can still be derived in this case by taking
velocity moments of the Vlasov equation, and identifying the velocity
dispersion with the above parameter $v_i$.
Thus although in the present paper we will refer to ``sound velocities'',
this should be taken as a generic expression for a velocity dispersion
parameter.
Such an approach should work fine for CDM, but may neglect an important
damping term found in HDM models.
The complete analogous equation to (\ref{general-cos-equs}) for HDM gives
a fluid equation perspective on free-streaming,
the phenomenon found in HDM models of neutrino-like matter.
An approximate equation has been derived for a hot neutrino-like component
by Setayeshgar \cite{setayeshgar}, which tends to wipe out perturbations
below a certain scale (see also the lecture notes by Bertschinger
\cite{bertschinger}).
In this work the exact Vlasov equation kinetic treatment was considered,
and the usual Fermi-Dirac distribution function was replaced by a
carefully chosen approximate form, allowing the conversion of the
integro-differential equation into the following pure differential
equation:
\be
\ddot{\delta}_{\nu} +
    \left( \frac{2\dot{a}}{a} + \frac{2\bar{k} v_{\nu}}{a^2} \right)
    \dot{\delta}_{\nu} + \frac{v_{\nu}^2 \bar{k}^{2}}{a^{2}} \delta_{\nu} = 
    4\pi G \sum_{i=1}^{n} \rho_{i}\delta_{i}.
\ee
Here the damping term $2\bar{k} v_{\nu}$ gives rise to
non-oscillating solutions heavily damped at short wavelengths.
Thus an equation of the form (\ref{general-cos-equs}) is not correct for
HDM of a neutrino-like nature.
We examine the general equations for both CDM and HDM, without specifying
too carefully the exact nature of the dark component involved.
This allows comparison of the results in this paper with previous work in
the literature, which has also neglected this point.
If some aspects of HDM models are poorly described by
(\ref{general-cos-equs}), the equations are still applicable to other
two-component cosmological systems such as a hydrogen-helium gas not in
equilibrium, where the lighter hydrogen component has a greater sound
speed.

At present (\ref{general-cos-equs}) has been displayed in a quite 
general form, with an unspecified scale factor $a$,
given by the Friedmann cosmological equation
\be
\frac{\dot{a}^{2}}{a^{2}} = \frac{8\pi G}{3}\rho + \frac{\Lambda}{3} -
    \frac{k_c}{a^{2}}.
\label{friedmann}
\ee
General parameters describing the nature of the Universe in this
equation are $k_c = 0, \pm 1$, the spatial curvature, and $\Lambda$, the 
cosmological constant.
It is difficult to make much progress without first becoming more
specific about the energy content of the Universe.
All the studies cited previously \cite{russians2,fargion,haubold1,haubold2} 
have only examined the Einstein-deSitter, matter dominated case, with 
various physical components and equations of state possible.
The studies \cite{haubold1,haubold2}, all of which are equivalent, make
the pretense to study the radiation dominated era as well, but this is
incorrect for the equations presented.
It was explicitly assumed that $a \propto t^{2/3}$ in the scaling of the
energy densities $\rho_{i} = \Omega_{i}/(6\pi Gt^{2})$ and the velocity 
parameters, yet an allowance was made for a general Hubble expansion
$H = \eta t^{-1}$.
The general parameter $\eta$ can only be equal to $\frac{2}{3}$ for the
equations presented to be physically correct.
The fluid equations were formulated to allow for general equations of state, 
by writing the sound velocities such that both their magnitudes and time
dependences were freely parameterized.
A range of solutions were obtained for different cases of the parameters,
and were generally classified by Meijer G-functions \cite{meijer,luke}.
We will show that the solutions found for a CDM and baryon model have
been evaluated incorrectly, and will proceed to find their general exact
representation.
We will also proceed to investigate a more general dark matter and
baryon problem than considered in any of the above.
In \cite{fargion}, several subcases of the above mentioned studies were
considered in some detail, and given a range of physical interpretations.
The solutions were of a mathematically simpler nature, involving either
Bessel functions or simple power law behavior.
The investigations in \cite{russians2} concentrated on a three-component
medium, involving baryons, CDM and photons.
They incorrectly used the nonrelativistic Newtonian cosmological equations
to model the photon component, so that the solutions, expressed in terms
of Meijer G-functions, cannot be considered as physically relevant.

Let us now make the choice of the matter dominated era of cosmological
evolution in which to set (\ref{general-cos-equs}), and in particular the
post-recombination era, where baryons had decoupled from photons.
This allows us to determine how the energy density and sound velocity
scale with respect to $a$, and consequently exhibit all explicit time
dependences in the equations.
We therefore introduce the comoving total background
density $\bar{\rho_{0}} \equiv a^{3} \rho_{0}$, and the constant
$\epsilon_{i} \equiv \rho_{i}/\rho_{0}$, the fraction of the total mass
density contributed by species $i$.
This is distinct from the in general time dependent quantity
$\Omega_{i}(t) \equiv \rho_{i}/\rho_{c}$, where $\rho_{c}$ is the critical 
density of the Universe
\be
\rho_{c} = \frac{3H^{2}}{8\pi G}.
\label{crit-density1}
\ee
We will consider a two-component fluid comprised of baryons (subscripted
by B) and dark matter (subscripted by D).
In the post-recombination era, the adiabatic speed of sound
of species $i$ assumes the following behavior:
\be
v_{i}^{2} \propto T_{i} \propto a^{-2},
\label{adiabatic-vel}
\ee
where $T_{i}$ is the temperature of the component.
This prompts us to define the time independent quantity 
$\bar{v}_{i}^2 \equiv a^{2}v_{i}^{2}$.
With these definitions, the linearized cosmological perturbation
equations may be written as
\ba
\label{2comp-gen(t)B}
\frac{d^{2}\delta_{B}}{dt^2} + 2\frac{\dot{a}}{a}\frac{d\delta_{B}}{dt}
    + \frac{\bar{v}_{B}^{2} \bar{k}^{2}}{a^{4}} \delta_{B} & = & 
    \frac{4\pi G\bar{\rho}_{0}}{a^{3}} (\eb\delta_B + \ed\delta_D), \\
\label{2comp-gen(t)D}
\frac{d^{2}\delta_{D}}{dt^2} + 2\frac{\dot{a}}{a}\frac{d\delta_{D}}{dt}
    + \frac{\bar{v}_{D}^{2} \bar{k}^{2}}{a^{4}} \delta_{D} & = & 
    \frac{4\pi G\bar{\rho}_{0}}{a^{3}} (\eb\delta_B + \ed\delta_D).
\ea

The equations currently still represent a fairly general cosmological
setting.
The curvature parameter $k_c$ and cosmological constant $\Lambda$ have
not been specified, and control the behavior of $a$ through the
Friedmann equation (\ref{friedmann}).
To see how these influence the evolution of the density perturbations, we
transform the dependent variable from $t$ to $a$.
We also use (\ref{friedmann}) and another cosmological
dynamics equation for the acceleration of $a$: 
\be
\ddot{a} = \frac{4}{3}\pi G \frac{\bar{\rho}_{0}}{a^{2}} + 
    \frac{\Lambda}{3}a.
\ee
This equation is derived in conjunction with the Friedmann equation by taking
the spatial components of the Einstein  equation.
The cosmological perturbation equations are now able to be written in
a form purely dependent on $a$, and parameterized explicitly by the
cosmological dynamical constants:
\ba
& (\frac{8}{3}\pi G\bar{\rho}_{0} + \frac{\Lambda}{3}a^{3} - k_c a)
    \delta_{B}^{''} + a^{-1}(4\pi G\bar{\rho}_{0} + \Lambda a^{3} - 2k_c a)
    \delta_{B}^{'} & \nonumber\\
\label{2comp-gen(a)B}
& + \frac{\bar{v}_{B}^{2} \bar{k}^{2}}{a^{3}} \delta_{B}
    - \frac{4\pi G\bar{\rho}_{0}}{a^{2}} 
    (\eb\delta_{B} + \ed\delta_{D}) = 0, & \\
& (\frac{8}{3}\pi G\bar{\rho}_{0} + \frac{\Lambda}{3}a^{3} - k_c a)
    \delta_{D}^{''} + a^{-1}(4\pi G\bar{\rho}_{0} + \Lambda a^{3} - 2k_c a)
    \delta_{D}^{'} & \nonumber\\
\label{2comp-gen(a)D}
& + \frac{\bar{v}_{D}^{2} \bar{k}^{2}}{a^{3}} \delta_{D}
    - \frac{4\pi G\bar{\rho}_{0}}{a^{2}} 
    (\eb\delta_{B} + \ed\delta_{D}) = 0. &
\ea
In the above, a prime denotes differentiation with respect to $a$.

A general analysis of (\ref{2comp-gen(a)B}) and (\ref{2comp-gen(a)D}) has
not been attempted previously.
To begin with, we can test for exactness of the equations (see 
\cite{murphy}, pp. 92, 93), to determine whether a first integral exists.
In general, a second order ordinary differential equation of the
form
\be
A_{0}(x)y^{''} + A_{1}(x)y^{'} + A_{2}(x)y = 0
\ee
[arbitrary functions $A_{i}(x)$] is exact if
\be
A_{0}^{''} - A_{1}^{'} + A_{2} = 0.
\ee
Let us first consider the one-component example.
This is the uncoupled case of (\ref{2comp-gen(a)B})
($\ed = 0,\; \eb = 1$), which gives
\be
A_{0}^{''} - A_{1}^{'} + A_{2} = \frac{\bar{v}_B^{2} \bar{k}^{2}}{a^{3}}.
\ee
Thus the one-component equation is only exact for a pressureless gas
$\bar{v}_{B} = 0$.
This means that no closed form solution is possible, and approximations 
need to be made.
We note that the pressureless one-component case has been studied
extensively (e.g. \cite{peebles}) for various values of the parameters.

To make progress with the perturbation equations, and also to make contact
with previous work in the literature, we need to
make some assumptions about $k_c$ and $\Lambda$.
The $k_c \neq 0$ cases tend to be more complicated mathematically, as
generally only parametric solutions can be found, where $a$ is represented
by hyperbolic functions ($k_c = -1$ open universe) or trigonometric 
functions ($k_c = 1$ closed universe).
Current observations, and the weight of theoretical tendencies in
cosmology (e.g. $\Omega = 1$ as demanded by inflation) make the choice of
flat universe $k_c = 0$ seem the most favorable.
$\Omega$ contains a contribution from $\Lambda$ as well as matter
components.
The large amount of observational data now being
analyzed, increasingly points to the existence of a cosmological constant
comprising a major fraction of the energy density (see 
e.g. \cite{perlmutter}, \cite{bennett}, \cite{perlmutter2}),
with a value of $\Omega_{\Lambda} \simeq 0.7$.
The Einstein-deSitter ($\Lambda = 0$, $\Omega = 1$) model is generally
not the model of choice anymore for detailed numerical studies in
cosmology, however we do not make a claim that the solutions
presented here are of an exact quantitative nature.
Many other factors must also be taken into account when attempting to build
up an exact, numerical model of structure formation.
We wish to correct and extend some previous results, as well as perform 
some new semi-quantitative analysis.
Our intent is to keep work analytically tractable at this stage.

We set $k_c = \Lambda = 0$ in (\ref{2comp-gen(a)B}) and (\ref{2comp-gen(a)D}).
In the Einstein-deSitter model, the critical density can be written
explicitly as
\be
\rho_{c} = \frac{\bar{\rho}_{0}}{a^{3}} = \frac{1}{6\pi Gt^{2}},
\label{crit-density2}
\ee
and the relation $\eb + \ed = 1$ holds.
We also introduce quantities resembling the comoving Jeans wavenumbers
for each component taken separately:
\be
\bar{k}_{B}^{2} = \frac{4\pi G\bar{\rho}_{0}}{\bar{v}_{B}^{2}},\;\;\;\;
    \bar{k}_{D}^{2} = \frac{4\pi G\bar{\rho}_{0}}{\bar{v}_{D}^{2}}.
\label{k-defs}
\ee
The difference with the true comoving Jeans wavenumber for a one-component
fluid is the inclusion of the total mass density $\bar{\rho}_0$, rather
than just the mass density of the component in question $\bar{\rho}_i$.
Equations (\ref{2comp-gen(a)B}) and (\ref{2comp-gen(a)D}) now become
\ba
\label{2compflatB}
\delta_{B}^{''} + \frac{3}{2a}\delta_{B}^{'} + \frac{3}{2a^{3}}
    \left( \frac{\bar{k}}{\bar{k}_{B}} \right)^{2} \delta_{B} & = &
    \frac{3}{2a^{2}}(\eb\delta_{B} + \ed\delta_{D}), \\
\label{2compflatD}
\delta_{D}^{''} + \frac{3}{2a}\delta_{D}^{'} + \frac{3}{2a^{3}}
    \left( \frac{\bar{k}}{\bar{k}_{D}} \right)^{2} \delta_{D} & = &
    \frac{3}{2a^{2}}(\eb\delta_{B} + \ed\delta_{D}).
\ea
The effects of the various physical processes are now clearly evident.
The expansion of the Universe  produces a damping term
$3\delta_{i}^{'}/(2a)$, causing the
solutions to be in power law form rather than exponential.
The relation of the mode wavenumber to the Jeans wavenumber is expressed
as a ratio, transparently showing in which region of physical scales the
mode lies.
This ratio can be compared to the fractions $\eb$ and $\ed$ to decide 
whether gravity or pressure dominates the dynamics.
The true Jeans instability scale for a two-component medium is not given by
either $\bar{k}_{B}$ or $\bar{k}_{D}$, but by a combination of the two, as
demonstrated in \cite{russians1,carvalho}.
This scale will be introduced in due course.

We finally perform a couple more manipulations, to cast the equations
in their simplest form.
We define the dimensionless parameters
\be
K_{B} = \frac{\bar{k}}{\bar{k}_{B}},\;\;\;\;
    K_{D} = \frac{\bar{k}}{\bar{k}_{D}}.
\ee
Then $K_{i} < 1$ corresponds to the Jeans unstable region in the
one-component analog of the equations, and $K_{i} > 1$ to the acoustic
region.
We also make the variable transformation $\chi = a^{-1/2}$.
This gives the final form of the system of differential equations to be
studied in the ensuing sections:
\ba
\label{canonicalB}
\delta_{B}^{''} + 6\left( K_{B}^2 - \frac{\eb}{\chi^2} \right) \delta_{B}
    - \frac{6\ed}{\chi^{2}}\delta_D & = & 0, \\
\label{canonicalD}
\delta_{D}^{''} + 6\left( K_{D}^2 - \frac{\ed}{\chi^2} \right) \delta_{D}
    - \frac{6\eb}{\chi^{2}}\delta_B & = & 0.
\ea
A prime now denotes differentiation with respect to $\chi$.
These equations bear a strong resemblance to the equations of an
electron-proton cosmological plasma studied in \cite{gailis} [equations
(4.8) and (4.9) of that paper].
As is well known from the analogy between the simple Jeans instability
and Langmuir modes, this resemblance is not surprising when the 
mathematical similarity between the electromagnetic and gravitational 
forces is considered.
The techniques employed in \cite{gailis} will be useful in our current
analysis.
In this paper we will employ the Frobenius method in obtaining long
wavelength solutions.
In a related paper \cite{paper2}, some general WKB techniques are
developed further than previously.
Apart from facilitating some short wavelength solutions to the current
problem, these techniques will also indicate further results 
possible in cosmological plasma physics.

Before we proceed to a general analysis of (\ref{canonicalB}) and
(\ref{canonicalD}), we wish to digress to the simpler case of a
one-component system.
Surprisingly, we will derive some apparently new results, which
provide a conceptually useful introduction to the ensuing analysis.

\section{The One-component Equation Revisited}

The Einstein-deSitter one-component equation for a baryonic or dark matter
fluid in the post-recombination era is a canonical example studied in 
all textbooks for linearized cosmological perturbation theory.
It gives the familiar Jeans unstable power law solutions 
$\delta \propto t^{2/3},\; t^{-1}$ in the limit of large scales, and
acoustic oscillations in the limit of small scales.
Despite this, we have not found the full exact solutions completely
displayed and analyzed in any textbooks or review articles in the current
literature.
Although a full analysis will not bring any startling new physical
revelations, the mathematical techniques required are of some interest
in their relation to the physics, and as an introduction to the more
complicated analysis we will require later.
This section may be seen as a useful orientation to the further work
carried out in the bulk of this paper.

We begin with the one-component version of (\ref{2compflatB}), i.e
with $\eb = 1$ and $\ed = 0$  [or vice versa for (\ref{2compflatD})].
Analogous to the definitions of $K_{B}$ and $K_{D}$, we define the
one-component comoving Jeans ratio for the fluid, which has comoving
Jeans wavenumber $\bar{k}_{J}$, as $K_{J} = \bar{k}/\bar{k}_{J}$.
The one-component density perturbation equation then becomes
\be
\delta^{''} + \frac{3}{2a}\delta^{'} + \left( \frac{3}{2a^{3}} K_{J}^{2}
    - \frac{3}{2a^{2}} \right) \delta = 0.
\label{onecomp-equ}
\ee
The solution of this equation is a Bessel function of order 5/2.
A Bessel function of half odd-integer order can be recast in terms of a
spherical Bessel function.
To begin with, we will choose the spherical Bessel functions of the
first and second kind, $j_{\nu}$ and $y_{\nu}$ respectively.
The solution may be rewritten as
\be
\delta(a) = c_{1} a^{-1/2} j_{2}\left( \sqrt{\frac{6}{a}} K_{J} \right)
    + c_{2} a^{-1/2} y_{2}\left( \sqrt{\frac{6}{a}} K_{J} \right),
\label{onecomp-solu}
\ee
with arbitrary constants of integration $c_{1}$ and $c_{2}$.
For this case the fortuitous circumstance arises that the solutions may
be represented in terms of elementary trigonometric functions (see e.g.
\cite{abram}).
The solutions as shown are exact mathematical representations, containing
all the information of the modes over all scales.
As is usually the case with such solutions, a simple inspection does
not reveal all the physical properties of the modes in an obvious manner.
For example, it is a little difficult to interpret the time dependence of
the modes through the argument of the Bessel functions $\sqrt{6/a} K_{J}$.
We will require various approximations and numerical plotting to extract
more physical meaning out of the solutions.

To begin with, we seek to place the solutions into a canonical form, for
easy comparison with other examples.
The most useful such form comprises, to leading order, a product of a 
power law time factor and complex exponential factor.
This approach was adopted in the studies of cosmological plasmas
\cite{dettmann,plasma,gailis} for one- and two-component systems.
As mentioned previously, due to the similarity between the gravitational
and electromagnetic forces, the corresponding modes display many 
similarities.

The most useful Bessel function solutions for our purposes are the Hankel
functions, due to the fact that their leading order terms contain complex
exponentials.
We use the spherical Hankel functions $h_{2}^{(1)}$ and $h_{2}^{(2)}$,
given by the expressions
\ba
\label{hankel1}
h_{2}^{(1)}(z) & = & \frac{1}{z} \exp\left[ i\left( z - \frac{3\pi}{2}
    \right) \right] \left(1 - \frac{3}{z^{2}} - \frac{3}{iz} \right), \\
\label{hankel2}
h_{2}^{(2)}(z) & = & \frac{1}{z} \exp\left[ -i\left( z - \frac{3\pi}{2}
    \right) \right] \left(1 - \frac{3}{z^{2}} + \frac{3}{iz} \right).
\ea
We may make the comparison here to plasma results, where analogous series
were obtained for large $z$.
Contrary to here, where the series has a finite number of terms, the 
series for plasma modes were only asymptotic.

We write down the explicit one-component solution via Hankel functions as
\be
\delta(a) = \left( 1 + \frac{a}{2K_{j}^{2}} + \frac{a^{2}}{4K_{J}^{4}}
    \right)^{1/2} \exp \left\{ \pm i \left[ \frac{\sqrt{6}K_{J}}{a^{1/2}}
    + \arctan \left( \frac{\sqrt{6}K_{J} a^{-1/2}}{2K_{J}^{2} a^{-1} -1}
    \right) \right] \right\}.
\label{onecomp-exp}
\ee
It must be stressed that unlike the plasma solutions, this is an exact
result.
The modulus of the solution grows with respect to time
to leading order as $\delta \propto a$ if $K_{J} \ll 1$, or else if 
$K_{J} \gg 1$ the modulus is approximately constant, with a first order
time correction proportional to $a$.
There is also a complex exponential portion to the solution, which
usually gives a dispersion relation.
The dispersion relation may be extracted by differentiating the phase 
with respect to $t$.
This follows from the general fact that given an observed frequency
$\omega$, a solution of the form
\be
\delta \propto \exp \left[ \pm i\int^{t} \omega(t)\,dt \right]
\ee
is expected.
This is assuming, of course, that the solution oscillates---if not, some 
other form of real valued solution must be available.
Using the matter dominated time dependence of $a$, namely
\be
a = \left( \frac{t}{t_{i}} \right)^{2/3},
\label{a-def}
\ee
where $t_{i}$ is an arbitrary constant, we find the frequency to be
\ba
\omega & = & \frac{\bar{v}_{s} \bar{k} a^{-2}}{1 + \frac{1}{2} K_{J}^{-2} a
    + \frac{1}{4} K_{J}^{-4} a^{2}} \nonumber\\
& \approx & \frac{\bar{v}_{s} \bar{k}}{a^{2}} \left( 1 - 
    \frac{1}{2} \frac{a}{K_{J}^{2}} + \frac{1}{8} \frac{a^{3}}{K_{J}^{6}} -
    \frac{1}{16} \frac{a^{4}}{K_{J}^{8}} + \cdots \right),\;\; K_{J} > 1.
\label{one-comp-omega}
\ea
The result has been expanded for $K_{J} > 1$, as we may suspect that due
to the Jeans instability, acoustic waves only exist in this region, and
thus we can only attach physical meaning to $\omega$ for $K_{J} > 1$.
This assertion will be derived rigorously in what ensues.

The result of (\ref{one-comp-omega}) may come as a surprise.
How does it relate to the well-known Jeans dispersion relation derived
for a static spacetime
\be
\omega^2 = v_{s}^{2} k^{2} - 4\pi G\rho_{0}\:?
\ee
In a cosmological setting, we may expect the dispersion relation to follow
a similar form, with appropriate time factors included.
For plasma modes, it was demonstrated in \cite{gailis} that the dispersion
relations could be written down to leading order in exactly the same form 
as their static spacetime counterparts in terms of physical (non-barred) 
variables, and then converted to comoving variables by inserting the 
correct time factors.
Thus we may expect
\be
\omega \approx \frac{\bar{v}_{s} \bar{k}}{a^{2}} 
    \left( 1 - \frac{a}{K_{J}^{2}} \right)^{1/2} 
\ee
at least in the form of a binomial expansion, namely
\be
\omega \sim \frac{\bar{v}_{s} \bar{k}}{a^{2}} \left( 1 - 
    \frac{1}{2} \frac{a}{K_{J}^{2}} - \frac{1}{8} \frac{a^{2}}{K_{J}^{4}}
    - \frac{1}{16} \frac{a^{3}}{K_{J}^{6}} + \cdots \right).
\label{postulate}
\ee
This form for $\omega$ may also be expected to contain some other time
dependent terms, as was demonstrated for a number of plasma modes.
Comparing the expansions in (\ref{one-comp-omega}) and (\ref{postulate}),
we see in fact that they only agree to first order.
This still indicates some form of Jeans instability, but the dispersion
relations are quite different.
This difference in behavior between the linearized gravitational modes
and plasma modes may be attributed to the special role the
density plays in the gravitational perturbation equations.
Eq.(\ref{onecomp-equ}) contains only one free 
parameter, the Jeans ratio $K_{J}$, whereas the plasma equations contain 
both the sound velocity and plasma frequency, which cannot be reduced to 
one parameter.
This implies that the relation between the gravitational source
and the Friedmann equation, which fixes the background spacetime, means
that the same form for the dispersion relation as found in static spacetime 
need not necessarily be expected in the expanding Einstein-deSitter model.

\begin{figure}
  \begin{center}
    \hspace{0cm}
    \epsfig{file=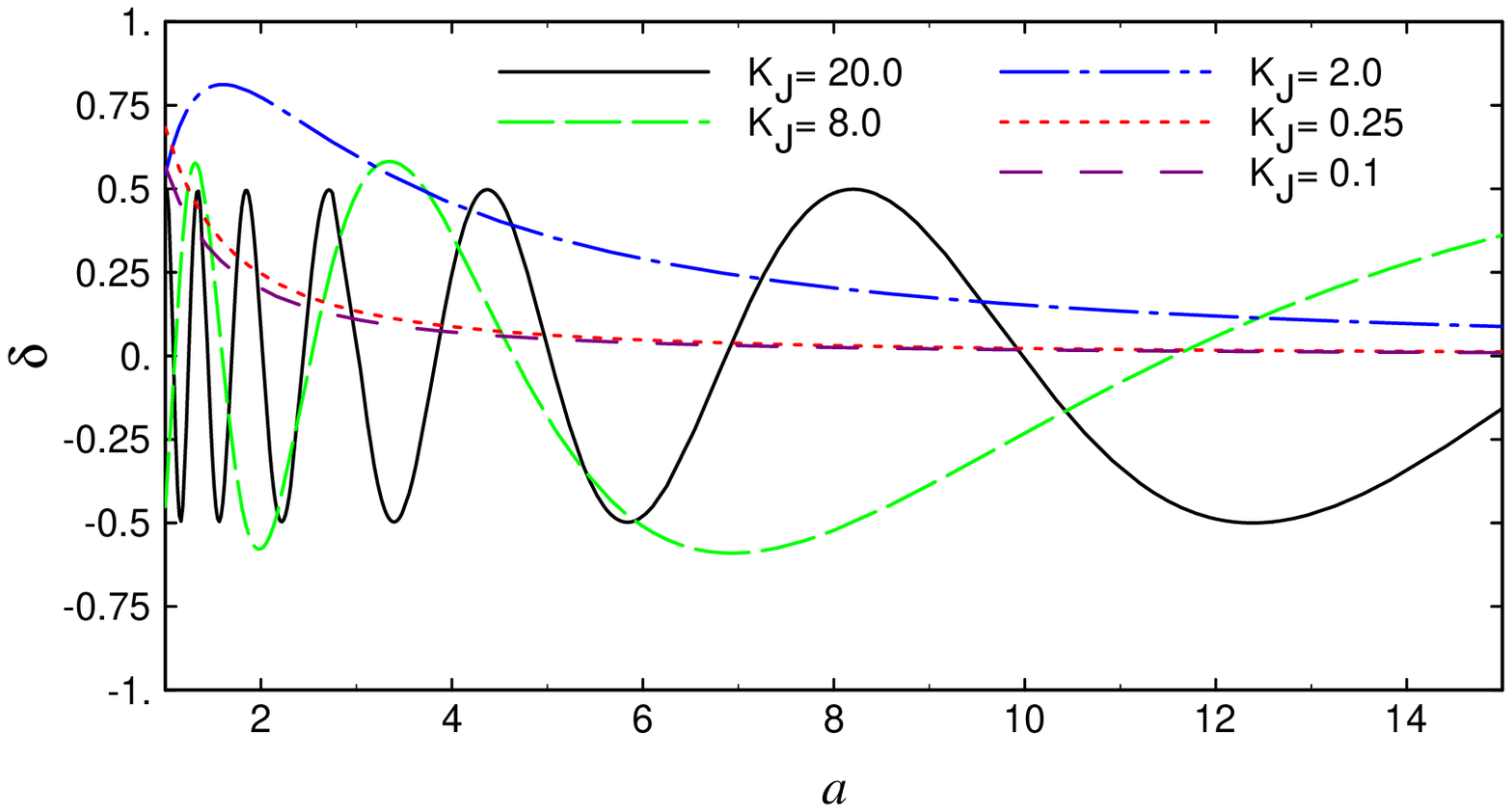, width=13cm}
  \end{center}
\caption{The transition of the decaying one-component modes as $K_J$
    varies through the Jeans instability}
\label{fig-decay}
\end{figure}

\begin{figure}
  \begin{center}
    \hspace{0cm}
    \epsfig{file=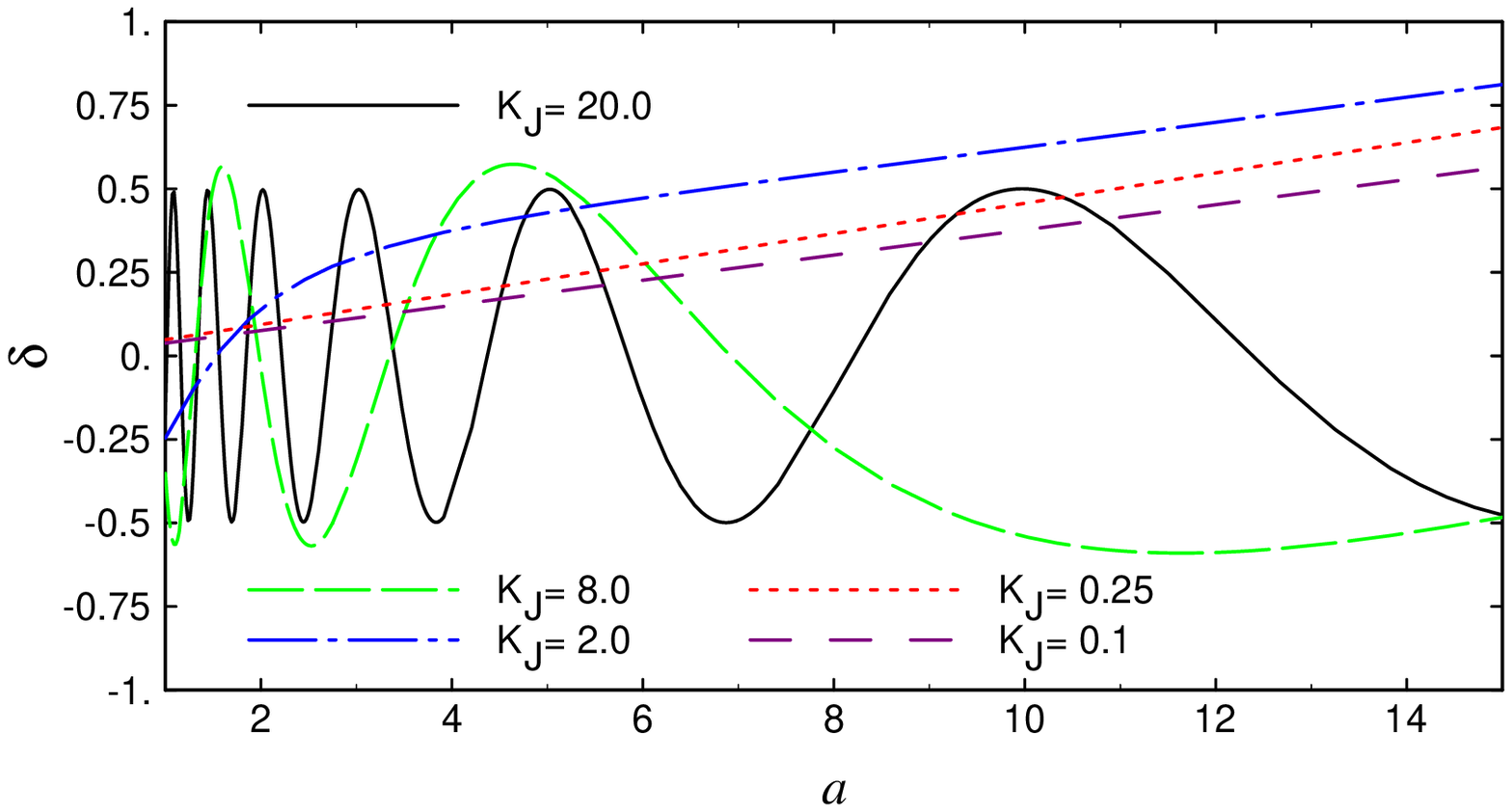, width=13cm}
  \end{center}
\caption{The transition of the growing one-component modes as $K_J$
    varies through the Jeans instability}
\label{fig-grow}
\end{figure}

We have a general solution in terms of a modulus and complex exponential,
which is exact and thus contains all the information of the problem.
How do we infer the usual Jeans instability behavior from this?
Let us examine plots of the solutions to gain a pictorial idea of what is
happening.
In Figs.~\ref{fig-decay} and \ref{fig-grow} we see the transition from
acoustic oscillations to growing and decaying modes as $K_{J}$ is
decreased---we are examining ever larger scales, and passing through the
instability.
To all fit on the same set of axes, the plots have been approximately
normalized.
The dependent variable $a$ is rather arbitrary [as can be deduced from
(\ref{a-def})].
An appropriate starting time $t_{i}$ may be chosen to normalize $a$ to 1
at the beginning of the chosen epoch of evolution, and the solutions may
then be propagated forward in time.
The tendency for the period of the acoustic oscillations to grow longer
in time is evident from the plots, and the approximate constancy of the
amplitude predicted previously from (\ref{onecomp-exp}) is evident.
In the extreme case, the oscillation period becomes so long that a
perturbation cannot complete one full oscillation, and then the instability
arises.
This behavior can clearly be seen in Figs.~\ref{fig-decay} and \ref{fig-grow}
for the $K_J = 2.0,\: 8.0$ plots.

We now perform some approximations to make contact with some better known
results of the one-component problem.
Let us begin with a small $K_{J}^{2}/a$ expansion.
For the spherical Hankel solutions (\ref{onecomp-exp}), we find
\be
\delta \sim \frac{a}{2K_{J}^{2}} \left[ 1 + \frac{K_{J}^{2}}{a} +
    O\left( \frac{K_{J}^{4}}{a^{2}} \right) \right] \exp \left\{
    \mp i\sqrt{6} \frac{K_{J}}{a^{1/2}} \left[ 1 - \frac{4}{5} 
    \frac{K_{J}^{4}}{a^{2}} + O\left( \frac{K_{J}^{6}}{a^{3}} \right)
    \right] \right\}.
\label{onecomp-smallKj}
\ee
The above expansion explains what happens to acoustic oscillations when
$K_{J} \lesssim 1$.
In this region, the leading order factor $K_{J} a^{-1/2}$ in the 
exponential must always lie between 1 and 0 numerically, and decreases with 
increasing time.
This is because $a \geq 1$ and increases monotonically for all time.
Thus the solution lies within one period of oscillation for all time,
and only the growing or decaying modes may be observed.
When $K_{J}$ becomes larger than 1, more than one period of oscillation may 
be spanned by the $K_{J} a^{-1/2}$ factor, and the solution will begin to
develop acoustic waves.
The expansion (\ref{onecomp-smallKj}) shows $\delta \propto a$, which only 
gives the familiar growing mode, discussed in all texts.
The decaying mode has not been found in the current analysis, because 
spherical Hankel functions have been chosen to represent the Bessel
function solutions.
The spherical Hankel functions are linear combinations
of the original $j_{2}$ and $y_{2}$ solutions, which contain both modes.
The decaying mode has been ``asymptotically swamped'' by the growing
mode in this linear combination.
The decaying mode may be liberated by a direct small variable expansion
of (\ref{onecomp-solu}) for each spherical Bessel function.
We find the $j_{2}$ component gives the decaying mode
\be
\delta \sim a^{-3/2} \left[ 1 - \frac{3}{7} \frac{K_{J}^{2}}{a} +
    O\left( \frac{K_{J}^{4}}{a^{2}} \right) \right],
\ee
and the $y_{2}$ component gives the same growing mode as 
(\ref{onecomp-smallKj}) in the slightly different form 
\be
\delta \sim a \left[ 1 + \frac{K_{J}^{2}}{a} + 
    O\left( \frac{K_{J}^{4}}{a^{2}} \right) \right]
\ee
without the exponential phase factor.
These solutions clearly correspond to the usual textbook modes found when
pressure is ignored.
They have included the pressure corrections, given as a series in the
Jeans ratio, with matching time factors to all orders in the expansion.

We now turn to the large parameter expansion, from which we expect to 
liberate the acoustic oscillations.
Eq.(\ref{onecomp-exp}) is in fact already in the form of a large
parameter expansion, only that it has a finite number of terms, and is 
consequently exact.
The phase of the exponential does not seem to show the structure of the
familiar Jeans dispersion relation.
This was discussed above, where it was indicated that this is not
necessarily to be expected.
It is possible to see why this is so in a lucid fashion by applying
the WKB method to the original equation (\ref{onecomp-equ}).
Through this method we will derive a dispersion relation displaying very
similar characteristics to the familiar textbook one, that shows the
presence of the Jeans instability.
A more complicated WKB approximation scheme was developed in \cite{gailis}
to deal with plasma modes of a more intricate form, involving larger
numbers of coupled equations.
This method is used in \cite{paper2} to handle the two-component 
cosmological density perturbation equations in the short wavelength
approximation.
For the present simple second order equation, the standard textbook 
approach suffices (for a good explanation of WKB methods, see 
\cite{bender}).

To see the physics most clearly, we transform (\ref{onecomp-equ}) to
depend on $t$.
Using the variable transformation given by (\ref{a-def}), the equation
\be
\ddot{\delta} + \frac{4}{3t} \dot{\delta} + \left( \frac{2}{3} K_{J}^{2}
    \frac{t_{i}^{2/3}}{t^{8/3}} - \frac{2}{3t^{2}} \right) \delta = 0
\label{t-equ}
\ee
results.
The usual Jeans dispersion relation can be directly seen in this equation.
Consider the factor $2/(3t^{2})$, arising from the gravitational source
term.
This term may be transformed to explicitly see the source 
parameters emerge.
We consider the physical (time dependent) value of the total energy density:
\be
4\pi G\rho_{0} = 4\pi G\rho_{c} = \frac{4\pi G}{6\pi Gt^{2}} = 
    \frac{2}{3t^{2}},
\ee
and the relation
\be
K_{J}^{2} = \frac{3}{2} \bar{v}_{s}^{2} \bar{k}^{2} t_{i}^{2}.
\ee
Then the Jeans dispersion relation can be directly seen in (\ref{t-equ}):
\be
\bar{v}_{s}^{2} \bar{k}^{2} \left( \frac{t_{i}}{t} \right)^{8/3} -
    \frac{2}{3t^{2}} = v_{s}^2 k^{2} - 4\pi G\rho_{0}.
\label{jeans-disp}
\ee
In the static spacetime case the physical variables would of course not
depend on time, and the first derivative term $4\dot{\delta}/(3t)$ in 
(\ref{t-equ}) would not exist.
This leads to the exact exponential solutions and the familiar dispersion
relation given by (\ref{jeans-disp}), as originally found by Jeans.

The most straightforward way to effect a WKB approximation in the present
situation is to remove the first derivative from the equation.
One way to do this is by the variable change $\chi = a^{-1/2}$.
We then find the equation
\be
\frac{d^2 \delta}{d\chi^2} + \left( 6K_J^2 - \frac{6}{\chi^2} \right)
    \delta = 0.
\label{WKBequ}
\ee
Now we suggestively define
\be
\tilde{\omega}(\chi) = \left( 6K_J^2 - \frac{6}{\chi^2} \right)^{1/2}.
\label{WKBomega}
\ee
Applying the WKB approximation to (\ref{WKBequ}) gives the leading order
solution
\be
\delta(\chi) \sim \tilde{\omega}^{-1/2} \exp \left[ \pm i\int
    \tilde{\omega}(\chi) d\chi \right].
\label{WKBsolu}
\ee
A dispersion relation has indeed been derived, and is given by
$\tilde{\omega}$ as defined in (\ref{WKBomega}).
If we consider
\ba
\int \tilde{\omega}(\chi) d\chi & = & -\int \left( \frac{2}{3} K_J^2 
    \frac{t^{2/3}}{t_i^{8/3}} - \frac{2}{3t^2} \right)^{1/2} dt \nonumber\\
& = & -\int \left( \frac{t}{t_{i}} \right)^{-4/3} \left[ 
    \bar{v}_{s}^{2}\bar{k}^{2} - \frac{2}{3t_{i}^{2}} 
    \left( \frac{t}{t_{i}} \right)^{2/3} \right]^{1/2} dt,
\ea
then a physical frequency $\omega(t)$ can be identified by use of 
(\ref{jeans-disp}) and the relation $t_{i}^{2} = 1/(6\pi G\bar{\rho}_{0})$.
Thus
\be
\int \tilde{\omega}(\chi) d\chi = \int \omega(t) dt = 
    \int (v_s^2 k^2 - 4\pi G\rho_{0})^{1/2} dt,
\ee
and the physical connection has been made.
Another point to note is that the amplitude $\tilde{\omega}(\chi)^{-1/2}$
is time independent to leading order (because $K_J \gg 1$).
Thus the amplitude is approximately constant, as noted previously.

To make a direct comparison of (\ref{WKBsolu}) and (\ref{onecomp-exp}),
we now evaluate the integral in the phase of the WKB solutions.
A change of integration variable from $\chi$ back to $a$ puts the integral
into a form which has been tabulated \cite{gradshteyn}, and we find
\be
\int \tilde{\omega}(\chi) d\chi = \left( \frac{6K_J^2}{a} - 6 \right)^{1/2} 
    - \frac{\sqrt{6}}{2} \arcsin \left( 1 - \frac{2a}{K_{J}^{2}} \right).
\ee
The solution generated by the WKB method may be compared to the exact
one given by (\ref{onecomp-exp}).
The amplitudes and phases need to be expanded for large $K_J$:
\ba
\lefteqn{\mbox{\rm WKB phase:} \hspace{5mm}
    \left( \frac{6K_J^2}{a} - 6 \right)^{1/2} -
    \frac{\sqrt{6}}{2} \arcsin \left( 1 - \frac{2a}{K_J^2} \right)} \nonumber\\
& & \hspace{4cm} = \sqrt{6} \frac{K_J}{a^{1/2}}
    \left[ 1 - \frac{\pi}{4} \frac{a^{1/2}}{K_J}
    + O \left( \frac{a}{K_J^2} \right) \right], \\
\lefteqn{\mbox{\rm exact phase:} \hspace{5mm}
    \frac{\sqrt{6}K_J}{a^{1/2}} + \arctan \left(
    \frac{\sqrt{6} K_J a^{-1/2}}{2 K_J^2 a^{-1} - 1} \right)}  \nonumber\\
& & \hspace{4cm} = \sqrt{6} \frac{K_J}{a^{1/2}} \left[ 1 + \frac{a}{2K_J^2} + 
    O \left( \frac{a}{K_J^2} \right) \right], \\
\lefteqn{\mbox{\rm WKB amplitude:} \hspace{5mm} (6K_J^2 - 6a)^{-1/4}
    \frac{1}{6^{1/4} K_J^{1/2}} \left[ 1 + \frac{a}{4K_J^2} +
    O \left( \frac{a^2}{K_J^4} \right) \right],} \\
\lefteqn{\mbox{\rm exact amplitude:} \hspace{5mm}
    \left( 1 + \frac{a}{2K_J^2} + \frac{a^2}{4K_J^4} \right)^{1/2}
    \left[ 1 + \frac{a}{4K_J^2} + O \left( \frac{a^2}{K_J^4} \right) \right].}
\ea
It can be seen that the two solutions only agree to leading order (modulo
time independent constant factors).
Given that the WKB method only gives a leading order solution to the
problem, no more can be expected.
This discussion has highlighted the difference between the expected
dispersion relations of static spacetime to those derived in an expanding
universe context.
WKB can reproduce the same form as the static spacetime dispersion relations,
but this may only agree to leading order to the true dispersion relation
found in an expanding universe scenario.

\section{Improvements on Previous CDM Perturbation Results}

We now return to the two-component equations and consider the case of
CDM perturbations characterized by strictly zero temperature.
If the velocity dispersion is considered to be an adiabatic sound velocity
as given by (\ref{adiabatic-vel}), then $T_i = 0$ corresponds to $K_D = 0$
in (\ref{canonicalD}).
Such an approximation facilitates an exact analytic solution to the
problem, which is otherwise impossible.
The resulting system of equations was one of the main cases investigated
in the general analysis of \cite{haubold2}.
In this section we will point out what appears to be an error in the
analysis of that paper, which leads to some markedly different solutions,
derived in what ensues.

Before we proceed with this, it is pertinent to point out a general
problem with taking $K_D = 0$ in Eqs.(\ref{canonicalB}),
(\ref{canonicalD}).
For simplicity, it is possible to neglect spacetime expansion, as the
qualitative behavior will be the same.
Thus the static spacetime cosmological equations studied in
\cite{carvalho} are sufficient for this discussion.
These equations are also examined in \cite{paper2}, and using the
notation employed there we have
\ba
\label{static-equD}
\ddot{\delta}_D + (v_D^2 k^2 - W_D) \delta_D - W_B \delta_B & = & 0, \\
\label{static-equB}
\ddot{\delta}_B + (v_B^2 k^2 - W_B) \delta_B - W_D \delta_D & = & 0,
\ea
with $W_i = 4\pi G\rho_i$.
This system of equations can be reduced to a first order linear autonomous
dynamical system describing a state vector
\be
{\bf x} = (x_1, x_2, x_3, x_4)^T \equiv
    (\dot{\delta}_D, \delta_D, \dot{\delta_B}, \delta_B)^T,
\ee
where $T$ denotes the transpose of a vector.
The dynamical system has a $k$-dependent critical point found by solving
the equation $\dot{\bf x} = 0$.
This critical point happens to give the Jeans wavenumber of the mixture,
and is given by
\be
k^2 = k_M^2 \equiv k_B^2 + k_D^2 = \frac{W_B}{v_B^2} + \frac{W_D}{v_D^2}.
\label{km-def}
\ee
Note the slight difference in this definition of $k_i$ in comparison to
$\bar{k}_i$ defined in (\ref{k-defs}).
It is already evident that a problem arises if we take $v_D \rightarrow 0$
in (\ref{km-def}), and this will be physically elucidated by studying
the four independent modes of the system, given by the eigenvalues of
the dynamical system.

The eigenvalues give the structure of the modes
(see \cite{carvalho} or \cite{paper2} for the details).
They are found to be of the general form
\be
\left\{
\begin{array}{ccccc}
    \lambda_{1} & = & -\lambda_{2} & = & \frac{1}{\sqrt{2}} \sqrt{f +
        \sqrt{f^{2} + 4g}} \\
    \lambda_{3} & = & -\lambda_{4} & = & \frac{1}{\sqrt{2}} \sqrt{f -
        \sqrt{f^{2} + 4g}}
\end{array}
\right. .
\label{eigenvalue}
\ee
For the case of CDM currently under consideration, where $v_D = 0$, the
$k$-dependent functions $f$ and $g$ are given by
\ba
\label{static-fdef}
f(k) & = & W_B + W_D - k^2 v_B^2, \\
\label{static-gdef}
g(k) & = & k^2 W_D^2 v_B^2.
\ea
It was previously found that the eigenvalues $\lambda_1$ and $\lambda_2$
described the Jeans unstable modes, whereas $\lambda_3$ and $\lambda_4$
described acoustic oscillations at all wavenumbers.
An examination of $\lambda_1$ in the current context will show this not
to be the case for CDM.
If we define (analogously to the one-component scenario)
\be
K_J = \frac{v_B^2 k^2}{W_B + W_D},
\ee
$\lambda_1$ may be written as follows:
\be
\lambda_1 = \frac{1}{\sqrt{2}} (W_B + W_D)^{1/2} \left\{ 1 - K_J^2 +
    \left[ (1 - K_J^2)^2 + 4\ed K_J^2 \right]^{1/2} \right\}^{1/2}.
\ee
This does not equal zero for $K_J = 1$, and it is straightforward to show
that it has no zeros for all $k \neq 0$.
Thus a CDM perturbation would collapse for {\em all} scales, clearly a
physical impossibility.

If we examine Eqs.(\ref{static-equD}), (\ref{static-equB}), it can
be seen that with the removal of the pressure term $v_D k^2$, there is
no mechanism to counter the remaining gravitational source terms
$-W_D \delta_D$ and $-W_B \delta_B$, whose sign indicate an
attractive forcing, initiating gravitational collapse.
This is the case no matter how small the fraction of dark matter compared
to baryons, thus no amount of baryonic pressure support can prevent a
collapse at any scale.
This physically absurd situation is the root of the problem of taking
$v_D \rightarrow 0$ in the fluid models.
A physically correct equation must include some sort of velocity dispersion
term, even if the matter is totally collisionless and an adiabatic speed
of sound cannot be defined.

With these thoughts in mind, we must view the current section as more of
a mathematical digression, than a physically realistic model.
It nevertheless serves a purpose.
We can make contact with the work of \cite{haubold2}, and uncover some
interesting mathematical properties associated with these cosmological
perturbation equations in general.
We find mathematical subtleties overlooked in \cite{haubold2}, which
also indicate the nature of the general solutions to follow in the next
section.
They allow an interesting comparison with cosmological plasma modes
discussed in \cite{gailis}, where the limit $T_D \rightarrow 0$ is valid.
The special nature of gravity, and the extra complications it entails are
revealed by this comparison.

We now proceed to obtain a solution of Eqs.(\ref{canonicalB}),
(\ref{canonicalD}) (with $K_D = 0$) in the long wavelength (small $k$)
limit.
A short wavelength solution is trivially obtained by setting $K_D = 0$
everywhere in the results presented in \cite{paper2}.
To align ourselves with earlier notation, and to stress the fact that
there is only one Jeans related scale now occurring, we will rename
$K_B$ to $K_J$.
Rather than directly reducing Eqs.(\ref{canonicalB}),(\ref{canonicalD}) into
a single equation, we attempt a solution by the Frobenius method.
This is useful as a precursor to the general solution derived in the
following section in a similar manner.

To begin with, we assume an arbitrary expansion of the solutions in the
form
\ba
\label{gen-frobB}
\delta_B (\rho, \chi) & = & \chi^{\rho} \sum_{n=0}^{\infty} a_n \chi^n, \\
\label{gen-frobD}
\delta_D (\rho, \chi) & = & \chi^{\rho} \sum_{n=0}^{\infty} b_n \chi^n.
\ea
Here $\rho$ is an arbitrary exponent to be determined, and $a_n$ and
$b_n$ are series coefficients also to be determined by the Frobenius
method.
We substitute these series into (\ref{canonicalB}) and
(\ref{canonicalD}) to obtain a set of algebraic relations
between the undetermined coefficients.
Then all coefficients of like power of $\chi$ are collected and equated to
zero.

Arising out of this procedure are a pair of indicial equations for $\rho$,
with one of $a_0$ or $b_0$ remaining an arbitrary constant:
\ba
\label{indicial1}
\rho (\rho - 1) a_0 - 6\eb a_0 - 6\ed b_0 & = & 0, \\
\label{indicial2}
\rho (\rho - 1) b_0 - 6\ed b_0 - 6\eb a_0 & = & 0.
\ea
A set of recursion relations also arise for all higher order coefficients:
\ba
\label{recursion1}
6K_J^2 a_n + [(\rho + n + 2)(\rho + n + 1) - 6\eb ] a_{n+2} -
    6\ed b_{n+2} & = & 0, \\
\label{recursion2}
(\rho + n + 2)(\rho + n + 1) - 6\ed ] b_{n+2} - 6\eb a_{n+2} & = & 0.
\ea
In retaining the arbitrary constant $a_0$ or $b_0$, it is assumed that all
odd indexed terms vanish from propagation of the initial values
$a_1 = b_1 = 0$ through the recursion relations.

If we solve (\ref{indicial1}) and (\ref{indicial2}) for $\rho$, we find
four possible values:
\be
\rho = 0,\: 1,\: 3,\: -2. \label{indicial-solu}
\ee
A comparison with the analogous case for cosmological plasma modes
\cite{gailis} immediately shows a difference in the nature of the exponents.
In the present case the exponents are exactly determined integers, whereas
for plasmas the exponents depended on the plasma frequency.
When the one-component solutions were discussed in the previous section,
an analogous difference was observed between the spherical Bessel function
solutions of the gravitational perturbation modes, and the general order
Bessel function solutions of the plasma modes.
Whereas in the one-component study the solutions were simplified by this
property of gravity, in the present case they are in fact complicated.
The plasma solutions were representable in terms of $_2 F_3$ generalized
hypergeometric functions, but a similar representation is not well-defined
here.
This is because the exponents differ by integers---a fact which
necessitates a modification of the basic Frobenius method.
This modification is borne out in the solutions by the fact that parameters
appearing in the denominator of generalized hypergeometric function
expansions cannot differ by integers.
In such situations the generalized hypergeometric functions are not
definable, and one must resort to classifying solutions of the equation
by Meijer G-functions.

To apply the Frobenius method to indices differing by integers, the
recursion relations must first be solved for general $\rho$.
This is achieved by employing (\ref{indicial1}), (\ref{indicial2}), and
writing the system of differential equations as
\be
{\bf L} \left[ \begin{array}{c}
                   \delta_B (\rho, \chi) \\ \delta_D (\rho, \chi)
               \end{array} \right] = 
\left[ \begin{array}{c}
           \rho (\rho - 1) a_0 - 6\eb a_0 - 6\ed b_0 \\
           \rho (\rho - 1) b_0 - 6\ed b_0 - 6\eb a_0
       \end{array} \right] \chi^{\rho-2},
\label{op-equ}
\ee
where the operator ${\bf L}$ is defined by
\be
{\bf L} = \left[ 
\begin{array}{cc}
    \ds{\frac{\partial^2}{\partial\chi^2} + 6\left( K_J^2 - 
        \frac{\eb}{\chi^2} \right)} &
    \ds{-\frac{6\ed}{\chi^2}} \\
    \ds{-\frac{6\eb}{\chi^2}} &
    \ds{\frac{\partial^2}{\partial\chi^2} - \frac{6\ed}{\chi^2}}
\end{array} \right].
\ee
At present there is no relation between $a_0$ and $b_0$, but the recursion
relations will provide one.
A lengthy algebraic exercise is needed to solve the recursion relations.
The result is
\ba
\label{coeffa}
a_{2n} & = & a_0 \,\frac{ (\frac{1}{2}\rho - \frac{1}{2}\nu_D + \frac{3}{4})_n
    (\frac{1}{2}\rho + \frac{1}{2}\nu_D + \frac{3}{4})_n }
    { (\frac{1}{2}\rho - \frac{1}{2})_n (\frac{1}{2}\rho + \frac{1}{2})_n
    (\frac{1}{2}\rho + 1)_n (\frac{1}{2}\rho + 2)_n }
    \left( \ts{-\frac{3}{2}} K_J^2 \right)^n, \\
\label{coeffb}
b_{2n} & = & \frac{6a_0 \eb}{\rho(\rho - 1) - 6\ed} \:
    \frac{ (\frac{1}{2}\rho - \frac{1}{2}\nu_D - \frac{1}{4})_n
    (\frac{1}{2}\rho + \frac{1}{2}\nu_D - \frac{1}{4})_n }
    { (\frac{1}{2}\rho - \frac{1}{2})_n (\frac{1}{2}\rho + \frac{1}{2})_n
    (\frac{1}{2}\rho + 1)_n (\frac{1}{2}\rho + 2)_n }
    \left( \ts{-\frac{3}{2}} K_J^2 \right)^n.
\ea
In the above, the parameter $\nu_D$ has been introduced as a shorthand:
\be
\nu_D \equiv \sqrt{\ts{\frac{1}{4}} + 6\ed},
\ee
and the notation $()_n$ is the Pochhammer symbol.
It is clear that the coefficients as derived will not exist for
$\rho = 1,\: -2$.
This is the basis for requiring a modification to the
straightforward method of substituting in the four calculated values of
$\rho$ (\ref{indicial-solu}) into (\ref{coeffa}) and (\ref{coeffb}) to
generate four independent solutions.

To proceed, we extract from (\ref{coeffa}) and (\ref{coeffb}) the relation
\be
b_0 = \frac{6a_0 \eb}{\rho(\rho - 1) - 6\ed},
\ee
which allows (\ref{op-equ}) to be rewritten as
\be
{\bf L} \left[ \begin{array}{c}
                   \delta_B (\rho, \chi) \\ \delta_D (\rho, \chi)
               \end{array} \right] = a_0
\left[ \begin{array}{c}
           \ds{\frac{1}{\eb} \frac{\rho(\rho - 1)(\rho + 2)(\rho - 3)}
           {\rho(\rho - 1) - 6\ed}} \\ 0
       \end{array} \right] \chi^{\rho-2}.
\ee
For $\rho = 0,\: 3$, direct substitution is permissible to obtain two
independent solutions, which turn out to be generalized hypergeometric
$_2 F_3$ functions, analogously to the plasma results.
For $\rho = 1,\: -2$, we take advantage of the fact that $a_0$ is arbitrary,
and set it respectively equal to $\rho - 1$ and $\rho + 2$.
After differentiation with respect to $\rho$ and evaluation at the
respective points $\rho = 1$ and $\rho = -2$, we obtain the result
\be
{\bf L} \left. \left[ 
\begin{array}{c}
    \ds{\frac{\partial}{\partial \rho} \delta_B (\rho, \chi)} \\ 
    \ds{\frac{\partial}{\partial \rho} \delta_D (\rho, \chi)}
\end{array} \right] \right|_{\rho = 1, -2} = 0.
\ee
This gives two more solutions.

The above discussion has defined an algorithm for finding small $k$
expansions of the solutions of the cosmological density perturbation
equations.
This algorithm is readily implemented into a symbolic manipulation
computer code.
We will write such a code to find the solutions to the general equations
in the next section.
The solutions for the CDM case under consideration may now be written
down.
The need to differentiate with respect to $\rho$ for the $\rho = 1,\: -2$
cases result in solutions involving digamma functions $\psi$ and 
logarithmic terms---quite a complication to the $_2 F_3$ functions
derived for the plasma perturbations in \cite{gailis}.
To display the modes in their simplest form, linear combinations of the
modes derived directly by the above algorithm need to be taken, and the
use of various mathematical identities involving digamma functions
employed.
The final set of four CDM modes are given as follows:
\ba
\label{cB1}
\delta_{B1} (\chi) & = & \frac{c_1}{\eb} \sum_{n=0}^{\infty}
    \frac{ (\frac{3}{4} - \frac{1}{2} \nu_D)_n (\frac{3}{4} +
    \frac{1}{2} \nu_D)_n } { (-\frac{1}{2})_n (\frac{1}{2})_n (2)_n n! }
    \left( -\ts{\frac{3}{2}} K_J^2 \chi^2 \right)^n \nonumber\\
& = & \frac{c_1}{\eb} \,_2 F_3 \left( \ts{\frac{3}{4} - \frac{1}{2} \nu_D,\:
    \frac{3}{4} + \frac{1}{2} \nu_D;\: -\frac{1}{2},\: \frac{1}{2},\: 2;\:
    -\frac{3}{2} K_J^2 \chi^2} \right) \\
\label{cD1}
\delta_{D1} (\chi) & = & \frac{c_1}{\ed} \sum_{n=0}^{\infty}
    \frac{ (-\frac{1}{4} - \frac{1}{2} \nu_D)_n (-\frac{1}{4} +
    \frac{1}{2} \nu_D)_n } { (-\frac{1}{2})_n (\frac{1}{2})_n (2)_n n! }
    \left( -\ts{\frac{3}{2}} K_J^2 \chi^2 \right)^n \nonumber\\
& = & \frac{c_1}{\ed} \,_2 F_3 \left( \ts{-\frac{1}{4} - \frac{1}{2} \nu_D,\:
    -\frac{1}{4} + \frac{1}{2} \nu_D;\: -\frac{1}{2},\: \frac{1}{2},\: 2;\:
    -\frac{3}{2} K_J^2 \chi^2} \right) \\
\label{cB2}
\delta_{B2} (\chi) & = & c_2 \frac{\chi}{\eb} + c_2 \frac{\chi}{\eb}
    \sum_{n=1}^{\infty} \frac{ (\frac{5}{4} - \frac{1}{2} \nu_D)_n 
    (\frac{5}{4} + \frac{1}{2} \nu_D)_n }
    { (\frac{3}{2})_n (\frac{5}{2})_n (n-1)! n! }
    \left( -\ts{\frac{3}{2}} K_J^2 \chi^2 \right)^n \nonumber\\
& & \mbox{} \times \left[ \ts{
    \psi \left( \frac{5}{4} - \frac{1}{2} \nu_D + n \right) +
    \psi \left( \frac{5}{4} + \frac{1}{2} \nu_D + n \right) -
    \psi (n) - \psi (n+1)} \right. \nonumber\\
& & \left. \mbox{} - \ts{\psi (n + \frac{3}{2}) - \psi (n + \frac{5}{2})
    + \log \chi^2} \right] \\
\label{cD2}
\delta_{D2} (\chi) & = & -c_2 \frac{\chi}{\ed} - c_2 \frac{\chi}{\ed}
    \sum_{n=1}^{\infty} \frac{ (\frac{1}{4} - \frac{1}{2} \nu_D)_n 
    (\frac{1}{4} + \frac{1}{2} \nu_D)_n }
    { (\frac{3}{2})_n (\frac{5}{2})_n (n-1)! n! }
    \left( -\ts{\frac{3}{2}} K_J^2 \chi^2 \right)^n \nonumber\\
& & \mbox{} \times \left[ \ts{
    \psi \left( \frac{1}{4} - \frac{1}{2} \nu_D + n \right) +
    \psi \left( \frac{1}{4} + \frac{1}{2} \nu_D + n \right) -
    \psi (n) - \psi (n+1)} \right. \nonumber\\
& & \left. \mbox{} - \ts{\psi (n + \frac{3}{2}) - \psi (n + \frac{5}{2})
    + \log \chi^2} \right] \\
\label{cB3}
\delta_{B3} (\chi) & = & c_3 \chi^3 \,_2 F_3 \left( \ts{
    \frac{9}{4} - \frac{1}{2} \nu_D,\: \frac{9}{4} + \frac{1}{2} \nu_D;\:
    \frac{5}{2},\: \frac{7}{2},\: 2;\: -\frac{3}{2} K_J^2 \chi^2} \right) \\
\label{cD3}
\delta_{D3} (\chi) & = & c_3 \chi^3 \,_2 F_3 \left( \ts{
    \frac{5}{4} - \frac{1}{2} \nu_D,\: \frac{5}{4} + \frac{1}{2} \nu_D;\:
    \frac{5}{2},\: \frac{7}{2},\: 2;\: -\frac{3}{2} K_J^2 \chi^2} \right) \\
\label{cB4}
\delta_{B4} (\chi) & = & c_4 \left( \ts{\frac{3}{2}} K_J^2 \chi^2
    \right)^{-1} + 2c_4 \ed \sum_{n=1}^{\infty}
    \frac{ (\frac{3}{4} - \frac{1}{2} \nu_D)_n 
    (\frac{3}{4} + \frac{1}{2} \nu_D)_n }
    { (-\frac{1}{2})_n (\frac{1}{2})_n (2)_n n! }
    \left( -\ts{\frac{3}{2}} K_J^2 \chi^2 \right)^n \nonumber\\
& & \mbox{} \times \left[ \ts{
    \psi \left( \frac{3}{4} - \frac{1}{2} \nu_D + n \right) +
    \psi \left( \frac{3}{4} + \frac{1}{2} \nu_D + n \right) -
    \psi (n+2) - \psi (n+1)} \right. \nonumber\\
& & \left. \mbox{} - \ts{\psi (n - \frac{1}{2}) -
    \psi (n + \frac{1}{2}) + \log \chi^2} \right] \\
\label{cD4}
\delta_{D4} (\chi) & = & c_4 \left( \ts{\frac{3}{2}} K_J^2 \chi^2
    \right)^{-1} - 2c_4 \eb \sum_{n=1}^{\infty} 
    \frac{ (-\frac{1}{4} - \frac{1}{2} \nu_D)_n 
    (-\frac{1}{4} + \frac{1}{2} \nu_D)_n }
    { (-\frac{1}{2})_n (\frac{1}{2})_n (2)_n n! }
    \left( -\ts{\frac{3}{2}} K_J^2 \chi^2 \right)^n \nonumber\\
& & \mbox{} \times \left[ \ts{
    \psi \left( -\frac{1}{4} - \frac{1}{2} \nu_D + n \right) +
    \psi \left( -\frac{1}{4} + \frac{1}{2} \nu_D + n \right) -
    \psi (n+2) - \psi (n+1)} \right. \nonumber\\
& & \left. \mbox{} - \ts{\psi (n - \frac{1}{2}) -
    \psi (n + \frac{1}{2}) + \log \chi^2} \right].
\ea

The same solutions may be obtained by considering Meijer G-function
solutions to the original differential equations.
The procedure for determining solutions to generalized hypergeometric-like
equations which contain parameters differing by integers is discussed in
detail by Luke \cite{luke}, pp.138-143.
The study outlined a method for developing Meijer G-function solutions
from the equations, which involved the differentiation of generalized
hypergeometric functions with respect to their parameters---a procedure
analogous to the differentiation of $\rho$ indices in the above.
The analysis involved is lengthy and tedious, but leads to the solutions
obtained above.
We refrain from a physical interpretation of the above gravitational modes
for now, and take that up in the next section when we discuss the more
general solutions.
The one obvious difference will be the fact that no Jeans instability is
apparent for any of the above modes, with two modes always collapsing and
two modes always acoustic, whereas for the general modes a Jeans instability
will be apparent.

We now compare the solutions obtained to those of \cite{haubold1,haubold2}.
In these studies general solutions were written as
\be
\phi_1 \equiv t^{-\alpha} \delta_B = c_1 G_1 + c_2 G_2 + c_3 G_3 + c_4 G_4.
\ee
Here $c_i$ are constants, and the functions $G_i$ denote
Meijer G-functions
\be
G_h = G_{2,4}^{m,n} \left( x \left|
\begin{array}{c}
    a_1^{*} + 1, a_2^{*} + 1 \\
    b_h^{*}, b_1^{*} \ldots \# \ldots b_4^{*}
\end{array} \right. \right),\;\;\; h = 1,\: 2,\: 3,\: 4.
\ee
The notation $\#$ signifies that $b_h$ is to be omitted in its usual place.
The G-function is defined so that $0 \leq m \leq 4$ and $0 \leq n \leq 2$.
The parameters $a_i^{*}, b_j^{*}$ depend on $\eb$, $\ed$, the adiabatic
index $\gamma_B$, and the exponent $\eta$ of $t$ in the Hubble expansion
parameter (which we pointed out earlier {\em must} be equal to
$\frac{2}{3}$ for the equations as formulated to be physically 
correct---even though \cite{haubold1,haubold2} used a greater range of
values).
The time parameter $x = \frac{3}{2} K_J^2 \chi^2$ in our notation.
A particular set of solutions is given by $m$ and $n$ being given specific
values.
In general, $m = 1$, $n = 2$ will give such a set of solutions for small
$x$ in the above example.
The G-functions can then normally be expressed in terms of $_2 F_3$
functions (for example the plasma solutions), but in the particular
case under discussion, since some of the $b_j^{*}$ differ by integers,
this is not possible.
It is this point that Haubold and Mathai missed in \cite{haubold2}.

Let us get down to specifics to illustrate the point.
Under the general classification scheme, the CDM case under consideration
corresponds to $\eta = \frac{2}{3}$, $\gamma_i = \frac{5}{3}$, in the
notation of \cite{haubold2}.
This implies that the parameters take the following values:
\ba
& a_1^{*},\: a_2^{*} = -1 \pm \frac{1}{2} \nu_D, & \nonumber\\
& b_1^{*},\: b_2^{*} = \pm \frac{1}{4},\;\; 
    b_3^{*},\: b_4^{*} = \pm \frac{5}{4}. & \nonumber
\ea
To evaluate the G-functions in this special case, where some parameters
differ by integers, the integral representation of the G-functions was
considered in \cite{haubold2}:
\be
G_1 = \frac{1}{2\pi i} \int_L 
    \frac{\Gamma(\frac{1}{4} + s) \Gamma(-a_1^{*} - s) \Gamma(-a_2^{*} - s)}
    {\Gamma(\frac{5}{4} - s) \Gamma(-\frac{1}{4} - s) \Gamma(\frac{9}{4} - s)}
    x^{-s} ds.
\ee
Using the fact that
\be
\frac{\Gamma(\frac{1}{4} + s)}{\Gamma(-\frac{1}{4} - s)}     -\frac{\Gamma(\frac{5}{4} + s)}{\Gamma(\frac{3}{4} - s)},
\ee
it was found that
\be
G_1 = -x^{5/4} \frac{\Gamma(-a_1^{*} + \frac{5}{4}) 
    \Gamma(-a_2^{*} + \frac{5}{4})}
    {\Gamma(\frac{5}{2}) \Gamma(2) \Gamma(\frac{7}{2})}
    \,_2 F_3 \left( -a_1^{*} + \frac{5}{4},\: -a_2^{*} + \frac{5}{4};\:
    \frac{5}{2},\: 2,\: \frac{7}{2};\: -x \right),
\ee
which agrees with $\delta_{B3}$ in (\ref{cB3}).
Performing the same analysis as above for $G_3$ however, it is found that
$G_1$ and $G_3$ are exactly the same function.
It appears as though this was never checked in \cite{haubold2}.
The same applies to $G_2$ and $G_4$:
\ba
G_2 = G_4 & = & -x^{-1/4} \frac{\Gamma(-a_1^{*} - \frac{1}{4}) 
    \Gamma(-a_2^{*} - \frac{1}{4})}
    {\Gamma(\frac{1}{2}) \Gamma(2) \Gamma(-\frac{1}{2})} \nonumber\\
& & \mbox{}\times \,_2 F_3 \left( -a_1^{*} - \frac{1}{4},\:
    -a_2^{*} - \frac{1}{4};\: \frac{1}{2},\: 2,\: -\frac{1}{2};\: -x \right),
\ea
which agrees with $\delta_{B1}$ in (\ref{cB1}).
Thus the solutions degenerate by employing this method, and two of the
solutions, namely $\delta_{B2}$ and $\delta_{B4}$ which contain
logarithmic terms, are totally missed.
The correct way to evaluate the G-functions when some parameters differ
by integers is given in Luke \cite{luke}, pp.143-147.
A careful study of this rather complicated procedure will show that the
logarithmic solutions $\delta_{B2}$ and $\delta_{B4}$ are found in this
way.

It is interesting to note that a linear combination of the solutions
$\delta_{B2}$ and $\delta_{B4}$ are in fact found in \cite{haubold2} in
the large $x$ limit.
In this case the function $G_{2,4}^{4,1}$ was evaluated from the contour
integral representation to achieve some analogous series to
(\ref{cB2}) and (\ref{cB4}).
Although the solutions (\ref{cB1})--(\ref{cD4}) are valid for all $x$,
such a representation does not seem very useful for $x$ large, as the
solutions are comprised of an infinite ascending series in $x$.
Thus very many terms would be required to represent the solutions 
accurately in this limit through $\delta_{B2}$ and $\delta_{B4}$.
In \cite{paper2} we develop a much better method for evaluating the
solutions for large $x$ using a WKB approximation scheme.

We have now thoroughly investigated the CDM two-component model using
the $K_D = 0$ limit, and tidied up previous work in this area.
As discussed earlier, although this work is of dubious physical relevance,
it has been an interesting mathematical investigation, and has allowed
comparisons with the previous work in cosmological perturbation theory
and cosmological plasma physics.
We now turn to the more general two-component model, which has a firmer
physical basis.

\section{The General Two-Component Solutions}

We finally investigate the most general set of equations, those with
both $K_B$ and $K_D$ nonzero.
The equations thus posed can model both CDM and HDM, though the discussion
initiated earlier about the lack of a free streaming damping term for
neutrino-like HDM should be heeded.

The algorithm needed to solve the equations was developed in the previous
section for $K_D = 0$, and can be used here.
Thus if (\ref{gen-frobB}) and (\ref{gen-frobD}) are substituted into
(\ref{canonicalB}) and (\ref{canonicalD}), we obtain the same indicial
equations (\ref{indicial1}) and (\ref{indicial2}) as previously, and the
following set of coupled recursion relations:
\ba
\label{gen-recursion1}
& (\rho + n + 2)(\rho + n + 1)(\rho + n + 4)(\rho + n - 1)
    a_{n+2} & \nonumber\\
& \mbox{} + 6K_B^2 (\rho + n + \frac{3}{2} - \nu_D)
    (\rho + n + \frac{3}{2} + \nu_D) a_n + 36 \ed K_D^2 b_n = 0, & \\
\label{gen-recursion2}
& (\rho + n + 2)(\rho + n + 1)(\rho + n + 4)(\rho + n - 1)
    b_{n+2} & \nonumber\\
& \mbox{} + 6K_D^2 (\rho + n + \frac{3}{2} - \nu_B)
    (\rho + n + \frac{3}{2} + \nu_B) b_n + 36 \eb K_B^2 a_n = 0. &
\ea
Here $\nu_B^2 = \frac{1}{4} + 6\epsilon_B$ in analogy with the previous
definition of $\nu_D$.
A closed solution for this set of recursion relations cannot be obtained
for all $n$, so that solutions to (\ref{canonicalB}) and (\ref{canonicalD})
can only be generated iteratively, to whatever order desired.
Lacking the ability to generate an infinite series representation for the
solutions means that they cannot be classified by known analytic functions.
To handle the complicated algebra involved in finding successive terms
iteratively, we have developed a symbolic computation code using the
functional programming language Mathematica.
The algorithm described in the previous section can be used to generate
a solution up to a certain power in $\chi$.

The coefficients increase in complexity very quickly for increasing $n$.
Although the code can generate solutions up to arbitrary order,
we find it sufficient to present only the first two orders for each
solution here.
We express the results in terms of the original parameters $K_B$ and
$K_D$, rather than in terms of $\nu_B$ and $\nu_D$, as no simplification
is gained in using the latter.
The solutions, corresponding to $\rho = 0,\: 1,\: 3,\: -2$ respectively,
are:
\ba
\label{gB1}
\delta_{B1} (\chi) & = & 1 + \frac{3}{2} (K_B^2 - 3\ed K_B^2 - 3 \eb K_D^2)
    \chi^2 + O(\chi^4), \\
\label{gD1}
\delta_{D1} (\chi) & = & -\frac{\eb}{\ed} \left[ 1 + \frac{3}{2}
    (K_D^2 - 3\ed K_B^2 - 3 \eb K_D^2) \chi^2 + O(\chi^4) \right], \\
\label{gB2}
\delta_{B2} (\chi) & = & \chi + \frac{1}{25} \left( 6K_B^2 - 31\ed K_B^2 -
    31\eb K_D^2 + 5\frac{\eb}{\ed} K_D^2 \right) \chi^3 \nonumber\\
& & \mbox{} + \frac{6}{5} \eb (K_D^2 - K_B^2) \chi^3 \log \chi
    + O(\chi^5), \\
\label{gD2}
\delta_{D2} (\chi) & = & -\frac{\eb}{\ed} \left[ \chi -
    \frac{1}{25} \left( 31\ed K_B^2 + 31\eb K_D^2 - K_D^2 \right) \chi^3
    \right. \nonumber\\
& & \left. \mbox{} - \frac{6}{5} \ed (K_D^2 - K_B^2) \chi^3 \log \chi
    + O(\chi^5) \right], \\
\label{gB3}
\delta_{B3} (\chi) & = & \chi^3 + \frac{3}{70} (3\ed K_B^2 - 3 \ed K_D^2
    - 10K_B^2) \chi^5 + O(\chi^7), \\
\label{gD3}
\delta_{D3} (\chi) & = & \chi^3 + \frac{3}{70} (3\eb K_D^2 - 3 \eb K_B^2
    - 10K_D^2) \chi^5 + O(\chi^7), \\
\label{gB4}
\delta_{B4} (\chi) & = & \chi^{-2} + \left( K_B^2 - 5\frac{\eb}{\ed} K_D^2
    + 5\ed K_B^2 - 5\ed K_D^2 \right) \nonumber\\
& & \mbox{} + 6\ed (K_B^2 - K_D^2) \log \chi + O(\chi^2), \\
\label{gD4}
\delta_{D4} (\chi) & = & \chi^{-2} + \left( 6K_D^2 - 5\eb K_B^2 + 5\eb K_D^2
    \right) \nonumber\\
& & \mbox{} + 6\ed (K_D^2 - K_B^2) \log \chi + O(\chi^2).
\ea

It remains now to make a physical interpretation of these solutions.
This is most usefully achieved by constructing a table comparing different
properties of the modes, and comparing the solutions to the corresponding
cosmological plasma modes.
The results are summarized in Table~\ref{table1}, which uses notation 
defined in \cite{gailis}.
Both the $K_D = 0$ modes of the previous section and the modes of this
current section are included in each category under the table.
The plasma modes $y_1 \ldots y_4$ of \cite{gailis} Eq.(4.16) correspond
to the $K_D = 0$ modes (\ref{cB1})--(\ref{cD4}).
The current gravitational modes (\ref{gB1})--(\ref{gD4}) (that is
$\delta_1 \ldots \delta_4$) correspond to the more general expansions
Eqs.(4.20), (4.21), (4.29)--(4.32) of \cite{gailis}.

\begin{table}
\begin{tabular}{cc} \hline
Gravitational Modes & Plasma Modes \\
\hline\hline
$\delta_1 \sim \chi^0 \sim t^0$ & $y_1 \sim \eta^0 \sim t^0$ \\
$\frac{\delta_B}{\delta_D} \sim -\frac{\ed}{\eb}$ & $\frac{n_e}{n_i} \sim 1$ \\
lower $_2 F_3$ parameters: $-\frac{1}{2},\: \frac{1}{2},\: 2$ &
  $\frac{1}{2},\: \frac{3}{4} - \frac{1}{2} \nu,\:
    \frac{3}{4} + \frac{1}{2} \nu$ \\
acoustic mode & ion-sound mode \\
\hline
$\delta_2 \sim \chi \sim a^{-1/2} \sim t^{-1/3}$ &
    $y_2 \sim \eta^{-1} \sim a^{-1/2} \sim t^{-1/3}$ \\
$\frac{\delta_B}{\delta_D} \sim -\frac{\ed}{\eb}$ & $\frac{n_e}{n_i} \sim 1$ \\
logarithmic solutions & parameters do not correspond \\
acoustic mode & ion-sound mode \\
\hline
$\delta_3 \sim \chi^3 \sim a^{-3/2} \sim t^{-1}$ &
    $y_3 \sim \eta^{-1/2-\nu} \sim a^{-1/4-(1/2)\nu} \sim t^{-1/6-(1/3)\nu}$ \\
$\frac{\delta_B}{\delta_D} \sim 1$ &
    $\frac{n_e}{n_i} \sim -\frac{P_e^2}{P_i^2}$ \\
lower $_2 F_3$ parameters: $-\frac{5}{2},\: \frac{7}{2},\: 2$ &
  $1 + \nu,\: \frac{3}{4} + \frac{1}{2} \nu,\:
    \frac{5}{4} + \frac{1}{2} \nu$ \\
collapsing mode & Langmuir mode \\
\hline
$\delta_4 \sim \chi^{-2} \sim a \sim t^{2/3}$ &
    $y_4 \sim \eta^{-1/2+\nu} \sim a^{-1/4+(1/2)\nu} \sim t^{-1/3+(1/3)\nu}$ \\
$\frac{\delta_B}{\delta_D} \sim 1$ &
    $\frac{n_e}{n_i} \sim -\frac{P_e^2}{P_i^2}$ \\
logarithmic solutions & parameters do not correspond \\
collapsing mode & Langmuir mode \\
\hline
\end{tabular}
\vspace{5mm}
\caption{A comparison of gravitational and plasma linear perturbation modes}
\label{table1}
\end{table}

In Table~\ref{table1}, the power of $\chi$ gives the corresponding
exponent $\rho$.
The parameter $\nu \equiv \sqrt{\frac{1}{4} - P_i^2 - P_e^2}$ found in
the plasma modes depends on the plasma frequencies of the electron and
ion components.
For the gravitational modes, the signs of $P_i$ and $P_e$ must be reversed
since (as opposed to the electromagnetic force) gravity is always
attractive.
In addition $P_i^2 + P_e^2$ corresponds to $\eb + \ed = 1$ (once
again the special nature of gravity in cosmology is apparent).
This implies that the general parameter $\nu$ in the plasma modes should
be replaced with $\frac{5}{2}$ for the gravitational modes---the
reason why many of the parameters in the $_2 F_3$ and Meijer G-functions
were pure rational numbers not depending on physical constants.
Notice that the ratio of the amplitudes of baryonic/dark matter modes
and electron/ion modes differ.
This is because the couplings in the differential equations are different.
For the gravitational modes the couplings involve terms such as
$\ed \delta_D$ and $\eb \delta_B$, whereas for the plasma modes the
couplings involve terms such as $P_i^2 \bar{n}_{e1}$ and
$P_e^2 \bar{n}_{i1}$.

We have indicated the corresponding collapsing and acoustic modes for
the gravitational density perturbations.
It is difficult to show this rigorously for the series solutions as
presented.
We can make comparisons to the one-component results, and identify the
leading order powers of the expansion parameter $a$.
This yields the classification as stated.
We can also make an analogy to the ion-sound modes of plasma physics,
which are of a similar nature to acoustic oscillations.
They show a collective behavior of both components oscillating approximately
in phase.

We are finally left with the question of how the Jeans scale enters into
the solutions.
In Section~4 the mixture wavenumber $k_M$, Eq.(\ref{km-def}) was briefly
introduced as being the only physically meaningful scale for instabilities
in a two-component fluid.
To make this quantity dimensionless, it would be appropriate to make the
definition
\be
K_M^2 = \frac{k^2}{W_B/v_B^2 + W_D/v_D^2}.
\ee
This quantity is only of relevance to a static spacetime scenario.
To place it in the context of the expanding Universe, the substitutions
\[
\begin{array}{ll}
W_B \rightarrow \ds{\frac{6\eb}{\chi^2}}, \hspace{1cm} &
    v_B^2 k^2 \rightarrow 6K_B^2, \\
W_D \rightarrow \ds{\frac{6\ed}{\chi^2}}, \hspace{1cm} &
    v_D^2 k^2 \rightarrow 6K_D^2
\end{array}
\]
are required.
Then $K_M$ takes on the revised definition
\be
K_M^2 = \frac{\chi^2}{\eb/K_B^2 + \ed/K_D^2} = \frac{\chi^2}{\chi_c^2}.
\ee
We have introduced the quantity $\chi_c(k)$, which can be thought of as a
critical time.
For $\chi > \chi_c$, $K_M > 1$ and acoustic oscillations would only be
expected to exist for all modes.
For $\chi < \chi_c$, $K_M < 1$ and two of the modes become unstable and
undergo gravitational collapse.
Since the precise magnitude of the scale factor $a$ is not determined by
cosmology [see Eq. (\ref{a-def}), which contains an arbitrary initial
time $t_i$], we may arbitrarily assign an initial time $a_0 = 1$, so that
$\chi_0 = 1$ and decreases with increasing time.
Then we may interpret the Jeans instability in two ways by considering
the critical time $\chi_c$.
Initially we may study all $k$-dependent modes at a particular time
$\chi$, where a subset will be unstable for values of $k$ for which
$\chi_c(k) > \chi$ (we stress that $\chi_c$ is a function of $k$).
We may then consider what occurs as the modes evolve through time from
this particular instant.
The critical time $\chi_c$ is fixed for any particular mode, so that a
subset of modes that were originally acoustic will become unstable as 
$\chi \rightarrow \chi_c^+$ (those modes corresponding to the solutions
$\delta_3$ and $\delta_4$ in Table I).
Consequently more and more modes pass through the instability as the
Universe evolves.
The physical wavenumber $k$ is of course dependent on time, thus the
dependence of the instability on a time $\chi_c$ shows the inextricable
link between the wavenumber and time.

It is illuminating at this stage to refer back to the one-component modes
discussed in Section~3.
For the one-component case, solutions were found in terms of the
combination $K_J a^{-1/2}$.
With the identification $\chi = a^{-1/2}$, the quantity $\chi_c$ is
seen to be the two-component analog of $K_J$.

It would be useful to convert the expansions (\ref{gB1})--(\ref{gD4}) to
depend on $K_M$ or $\chi_c$, to see how the Jeans scale enters.
This is achieved by the following relations:
\be
K_B^2 = \frac{1}{\chi_c^2}(\eb + \ed V^2), \hspace{1cm}
    K_D^2 = \frac{1}{\chi_c^2}(\ed + \frac{\eb}{V^2}).
\ee
Here $V = v_B/v_D$ is the ratio of sound velocities.
We then find that the expansions are all in terms of increasing powers
of $\chi/\chi_c$, with coefficients in terms of $\eb$, $\ed$ and $V$:
\ba
\delta_{B1}(\chi) & = & 1 + \frac{3}{2} \left( \eb + \ed V^2 - 6\eb\ed
    - 3\ed^2 V^2 - \frac{3\eb^2}{V^2} \right) \frac{\chi^2}{\chi_c^2}
    + O \left( \frac{\chi^4}{\chi_c^4} \right), \\
\delta_{D1}(\chi) & = & -\frac{\eb}{\ed} \left[ 1 + \frac{3}{2}
    \left( \ed + \frac{\eb}{V^2} - 6\eb\ed - 3\ed^2 V^2 - 
    \frac{3\eb^2}{V^2}\right) \frac{\chi^2}{\chi_c^2} \right. \nonumber\\
& & \left. \mbox{} + O \left( \frac{\chi^4}{\chi_c^4} \right) \right] \\
\delta_{B2}(\chi) & = & \chi \left\{ 1 + \frac{1}{25} \left[
    5 + \ed + \frac{6\eb}{V^2} + \frac{5\eb^2}{\ed V^2} -
    31 \left( \ed V + \frac{\eb}{V} \right)^2 \right]
    \frac{\chi^2}{\chi_c^2} \right. \nonumber\\
& & \left. \mbox{} + \frac{6}{5} \eb \left( \ed - \eb - \ed V^2 +
    \frac{\eb}{V^2} \right) \frac{\chi^2}{\chi_c^2} \log \chi +
    O \left( \frac{\chi^4}{\chi_c^4} \right) \right\}, \\
\delta_{D2}(\chi) & = & -\frac{\eb}{\ed}\chi \left\{ 1 - \frac{1}{25}
    \left[ \ed - \frac{\eb}{V^2} + 31 \left( \ed V + \frac{\eb}{V} \right)^2
    \right] \frac{\chi^2}{\chi_c^2} \right. \nonumber\\
& & \left. \mbox{} - \frac{6}{5} \ed \left( \ed - \eb - \ed V^2 +
    \frac{\eb}{V^2} \right) \frac{\chi^2}{\chi_c^2} \log \chi +
    O \left( \frac{\chi^4}{\chi_c^4} \right) \right\}, \\
\delta_{B3}(\chi) & = & \chi^3 \left\{ 1 + \frac{3}{70} \left[ 3\ed \left(
    \eb - \ed + \ed V^2 - \frac{\eb}{V^2} \right) \right.\right. \nonumber\\
& & \left.\left. \mbox{} - 10(\eb + \ed V^2) \right] \frac{\chi^2}{\chi_c^2}
    + O \left( \frac{\chi^4}{\chi_c^4} \right) \right\}, \\
\delta_{D3}(\chi) & = & \chi^3 \left\{ 1 + \frac{3}{70} \left[ 3\eb \left(
    \ed - \eb - \ed V^2 + \frac{\eb}{V^2} \right) \right.\right. \nonumber\\
& & \left.\left. \mbox{} - 10 \left( \ed + \frac{\eb}{V^2} \right) \right]
    \frac{\chi^2}{\chi_c^2} + O \left( \frac{\chi^4}{\chi_c^4} \right)
    \right\},
\ea
\ba
\delta_{B4}(\chi) & = & \chi^{-2} \left\{ 1 + \left[ -4\eb + \ed V^2 +
    5\eb\ed + 5\ed^2 (V^2 - 1) \right.\right. \nonumber\\
& & \left.\left. \mbox{} - 5\left( \frac{\eb}{\ed} + \ed \right)
    \frac{\eb}{V^2} \right] \frac{\chi^2}{\chi_c^2} \right. \nonumber\\
& & \left. \mbox{} + 6\ed \left( \eb - \ed + \ed V^2 - \frac{\eb}{V^2}
    \right) \frac{\chi^2}{\chi_c^2} \log \chi +
    O \left( \frac{\chi^4}{\chi_c^4} \right) \right\}, \\
\delta_{D4}(\chi) & = & \chi^{-2} \left\{ 1 + \left[ 6\ed - 5\eb^2 +
    5\eb\ed (1 - V^2) + \frac{5\eb^2}{V^2} \right] \frac{\chi^2}{\chi_c^2} 
    \right. \nonumber\\
& & \left. \mbox{} - 6\eb \left( \eb - \ed + \ed V^2 - \frac{\eb}{V^2}
    \right) \frac{\chi^2}{\chi_c^2} \log \chi +
    O \left( \frac{\chi^4}{\chi_c^4} \right) \right\}.
\ea
This is a convenient parameterization of the solutions.
The scale of the modes are chosen by $\chi_c$, the nature of the matter
involved is determined by $V$, and the proportions are determined by
$\eb$ and $\ed$.
A complete solution to the problem has thus been achieved up to whatever
order desired.

\section{Conclusions and Further Work}

A method for determining the small $k$ solutions of a general two-component
cosmological density perturbation model has been expounded in this paper.
We have only displayed the solutions to first order, but it is possible
derive them up to any order by the method in principle.
We have explored the mathematical properties and peculiarities of density
perturbations influenced by gravitational interaction, particularly
contrasting them to plasma modes, and correcting a number of previous
misconceptions in the literature.
The expanding Universe introduces new features not predictable from
simple static spacetime considerations.
In particular, totally new structures to the dispersion relations are
found, even in the one-component example.
We have shown how the mixture Jeans wavenumber enters the solutions, and
clarified its role in an expanding universe context.

More work is required to investigate the solutions around the critical
scale defined by $k_M$.
Although the expansions as derived in this paper are applicable to this
region, they are not particularly useful, as many terms in the equations
need to be retained when the expansion parameter $\chi/\chi_c$ is of
$O(1)$.
It is unclear how an analytical investigation of this region could proceed
at present.
We have performed some preliminary studies which involved producing a
large number of terms in the expansions (\ref{gB1})--(\ref{gD4}) using
the Mathematica program described, and then substituting in numerical
values for the various physical parameters to obtain numerical
coefficients with an ascending series in $\chi$.
At present the plots of these expansions over a range of values of $\chi$
do not yield reliable results---it is possible that many more terms than
are practically calculable will be required, and a very high order of
numerical precision will have to be maintained.
Other methods of analyzing the modes in this interesting region probably
need to be investigated.

Of ultimate interest is exploring how these type of modes contribute to
the power spectrum.
More physical effects may need to be introduced, such as a cosmological
constant, or the addition of more matter components.
To determine the actual density contrast at a given scale $1/k$, the
Fourier modes of the density contrast as derived in this paper would
also need to be integrated over the whole range $0 < k < 1/k$.
It would be of considerable interest to compare the power spectra
calculated by such a method with the well-known power spectra of the
various cosmological models in existence today.

In concluding, we remark that a similar analysis could be carried out in
the post-recombination region $140 < z < 1150$, where now the baryons follow
the $T \sim 1/a$ relationship.
The differential equations in Section II will now be different, as will be
their solutions; but we expect
that the ensuing analysis would yield qualitatively similar results but
quantitatively different scaling.
This would be a useful future study.

\section*{Acknowledgements}

The authors would like to thank the Australian Research Council for
funding this work.
We are grateful to Neil Cornish for pointing out the physics given at the
end of the Introduction.

\newpage

\begin{thebibliography}{00}
 
\bibitem{padman} T. Padmanabhan, {\em Structure Formation in the Universe}
    (Cambridge University Press, Great Britain, 1993).
\bibitem{peebles} P. J. E. Peebles, {\em The Large Scale Structure of
    the Universe} (Princeton University Press, Princeton, 1980);
    P. J. E. Peebles, {\em Principles of Physical Cosmology} (Princeton
    University Press, Princeton, 1993).
\bibitem{virgo} A. Jenkins, C. S. Frenk, F. R. Pearce, P. A. Thomas,
    J. M. Colberg, S. D. M. White, H. M. P. Couchman, J. A. Peacock,
    G. Efstathiou, and A. H. Nelson, Astrophys. J. {\bf 499}, 20 (1998).
\bibitem{texts} S. Weinberg, {\em Gravitation and Cosmology} (Wiley,
    New York, 1972); E. W. Kolb and M. S. Turner, {\em The Early Universe}
    (Addison-Wesley, New York, 1994); Ya. B. Zel'dovich and I. D. Novikov,
    {\em The Structure and Evolution of the Universe, Relativistic
    Astrophysics} (University of Chicago Press, Chicago, 1983), Vol 2.
\bibitem{jeans} J. H. Jeans, Phil. Trans. R. Soc. Long. A {\bf 199}, 
    1 (1902); J. H. Jeans, {\em Astronomy and Cosmology}, (Cambridge
    University Press, Cambridge, Great Britain, 1929).
\bibitem{russians1} V. L. Polyachenko and A. M. Fridman, Sov. Phys. JETP
    {\bf 54}, 7 (1981); L. P. Grishchuk and Ya. B. Zel'dovich, Sov. Astron.
    {\bf 25}, 267 (1981).
\bibitem{carvalho} J. P. M. de Carvalho and P. G. Macedo, Astron. Astrophys.
    {\bf 299}, 326 (1995).
\bibitem{russians2} L. V. Solov'eva and I. S. Nurgaliev, Sov. Astron. 
    {\bf 29}, 267 (1985); L. V. Solov'eva and A. A. Starobinsky, Sov. Astron.
    {\bf 29}, 367 (1985); I. S. Nurgaliev, Sov. Astron. Lett. {\bf 12},
    73 (1986).
\bibitem{fargion} D. Fargion, Nuovo Cim. {\bf B77}, 111 (1983).
\bibitem{haubold1} A. M. Mathai, H. J. Haubold, J. P. M\"{u}cket, 
    S. Gottl\"{o}ber, and V. M\"{u}ller, J. Math. Phys. {\bf 29}, 2069 (1988);
    A. M. Mathai, Studies Appl. Math. {\bf 80}, 75 (1989); H. J. Haubold, 
    A. M. Mathai, and J. P. M\"{u}cket, Astron. Nachr. {\bf 312}, 1 (1991).
\bibitem{haubold2} H. J. Haubold and A. M. Mathai, Astrophys. and Space Sci.
    {\bf 214}, 139 (1994).
\bibitem{lifshitz} E. M. Lifshitz, J. Phys. (USSR) {\bf 10}, 116 (1946); 
    E. M. Lifshitz, Zh. Eksp. Teoret. Fiz. {\bf 16}, 587 (1947).
\bibitem{novikov} I. D. Novikov, Sov. Phys. JETP {\bf 19}, 467 (1964).
\bibitem{ratra} B. Ratra, Phys. Rev. D {\bf 37}, 3406 (1988); B.Ratra and
    P. J. E. Peebles, Phys. Rev. D {\bf 52}, 1837 (1995).
\bibitem{buchert} S. Adler and T. Buchert, Astron. Astrophys. {\bf 343}, 317
    (1999); T. Buchert, A. Dominguez and J. Perez-Mercader, Astron. Astrophys.
    {\bf 349}, 343 (1999); T. Buchert and A. Dominguez, Astron. Astrophys.
    {\bf 438}, 443 (2005).
\bibitem{dettmann} C. P. Dettmann, N. E. Frankel and V. Kowalenko, 
    Phys. Rev. D {\bf 48}, 5655 (1993).
\bibitem{plasma} R. M. Gailis, C. P. Dettmann, N. E. Frankel and 
    V. Kowalenko, Phys. Rev. D {\bf 50}, 3847 (1994); R. M. Gailis, 
    N. E. Frankel and C. P. Dettmann, Phys. Rev. D {\bf 52}, 6901 (1995).
\bibitem{gailis} R. M. Gailis and N. E. Frankel, Phys. Rev. D {\bf 56},
    7750 (1997).
\bibitem{paper2} R. M. Gailis and N. E. Frankel, {\em Short Wavelength
    Analysis of the Evolution of Perturbations in a Two-component
    Cosmological Fluid}, submitted to J. Math. Phys. the following paper.
\bibitem{setayeshgar} S. Setayeshgar, SB thesis (MIT, 1990).
\bibitem{bertschinger} E. Bertschinger, ``Cosmological Dynamics'' in
    {\em 1993 Les Houches Summer School Lectures on Cosmology} (Elsevier
    Science Publishers, 1995).
\bibitem{meijer} C. S. Meijer, Proc. Nederl. Akad. Wetensch. {\bf A49},
    344 (1946).
\bibitem{luke} Y. L. Luke, {\em The Special Functions and their
    Approximations} (Academic, New York, 1969), Vol. 1.
\bibitem{murphy} G. M. Murphy, {\em Ordinary Differential Equations and
    their Solutions}, (D. Van Nostrand Company, Princeton, New Jersey, 1960).
\bibitem{perlmutter} S. Perlmutter {\em et al}, Astrophys. J. {\bf 517}
  565 (1999).
\bibitem{bennett} C. L. Bennett {\em et al}, Astrophys. J. Suppl. {\bf 148},
  1 (2003).
\bibitem{perlmutter2} S. Perlmutter and B. Schmidt, ``Measuring Cosmology
    and Supernovae'', in {\em Supernovae and Gamma Ray Bursts (Lecture Notes
    in Physics)}, K. Weiler, ed., (Springer--Verlag Berlin Heidelberg, 2003).
\bibitem{abram} M. Abramowitz and I. A. Stegun, {\em Handbook of Mathematical
    Functions}, (Dover, New York, 1964).
\bibitem{bender} C. M. Bender and S. A. Orszag, {\em Advanced Mathematical
    Methods for Scientist and Engineers} (McGraw-Hill, Singapore, 1978).
\bibitem{gradshteyn} I. S. Gradshteyn and I. M. Ryzhik, {\em Table of
    Integrals, Series and Products} (Academic Press, New York, 1965).

\end{thebibliography}
\end{document}